\documentclass[11pt,a4paper]{report}
\usepackage[utf8]{inputenc}
\usepackage[T1]{fontenc}
\usepackage{geometry}
\usepackage{fancyhdr}
\usepackage{titlesec}
\usepackage{enumitem}
\usepackage{hyperref}
\usepackage{parskip}

\title{Recommendations for Best Practices for Data Preservation and Open Science in HEP}

\author{Anne Gentil-Beccot$^{1}$, Antonia Winkler$^{2,1}$, Caterina Doglioni$^{3}$ \\
Christoph Wissing$^{4}$, Clemens Lange$^{5}$, Cristinel Diaconu$^{6,7}$ \\
Dillon S. Fitzgerald$^{8}$, Elizabeth Sexton-Kennedy$^{9}$, Eric Lancon$^{10}$ \\
Fazhi Qi$^{11}$, Fleur Heiniger$^{1}$, Gang Chen$^{11}$ \\
Giacomo Tenaglia$^{1}$, Graeme Andrew Stewart$^{4}$, Gustavo Valdiviesso$^{12}$ \\
Hao Hu$^{11}$, Harvey Newman$^{13}$, Ianna Osborne$^{14}$ \\
Irakli Chakaberia$^{15}$, Julie M. Hogan$^{16}$, Kati Lassila-Perini$^{17}$$^{*}$ \\
Michael D. Hildreth$^{18}$, Michael Sparks$^{3}$, Mihoko Nojiri$^{19}$ \\
Nicola Tarocco$^{1}$, Olivia Mandica-Hart$^{1}$, Salomé Rohr$^{1}$ \\
Seema Sharma$^{20}$, Simone Campana$^{1}$, Stefan Roiser$^{1}$ \\
Thomas McCauley$^{18}$, Thomas Schörner$^{4}$, Tibor Šimko$^{1}$ \\
Ulrich Schwickerath$^{1}$, Vincent Garonne$^{10}$, Zach Marshall$^{15}$}
\date{\today}

\begin{document}

\begin{titlepage}
\begin{center}
\vspace{1cm}
{\LARGE \textbf{Recommendations for Best Practices for Data Preservation\\and Open Science in HEP} \par}
\vspace{1cm}
{\normalsize
Simone Campana$^{1}$, Irakli Chakaberia$^{2}$, Gang Chen$^{3}$ \\
Cristinel Diaconu$^{4,5}$, Caterina Doglioni$^{6}$, Dillon S. Fitzgerald$^{7}$ \\
Vincent Garonne$^{8}$, Anne Gentil-Beccot$^{1}$, Fleur Heiniger$^{1}$ \\
Michael D. Hildreth$^{9}$, Julie M. Hogan$^{10}$, Hao Hu$^{3}$ \\
Eric Lancon$^{8}$, Clemens Lange$^{11}$, Kati Lassila-Perini$^{12}$$^{*}$ \\
Olivia Mandica-Hart$^{1}$, Zach Marshall$^{2}$, Thomas McCauley$^{9}$ \\
Harvey Newman$^{13}$, Mihoko Nojiri$^{14}$, Ianna Osborne$^{15}$ \\
Fazhi Qi$^{3}$, Salomé Rohr$^{1}$, Stefan Roiser$^{1}$ \\
Thomas Schörner$^{16}$, Ulrich Schwickerath$^{1}$, Elizabeth Sexton-Kennedy$^{17}$ \\
Seema Sharma$^{18}$, Tibor Šimko$^{1}$, Michael Sparks$^{6}$ \\
Graeme Andrew Stewart$^{16}$, Nicola Tarocco$^{1}$, Giacomo Tenaglia$^{1}$ \\
Gustavo Valdiviesso$^{19}$, Antonia Winkler$^{20,1}$, Christoph Wissing$^{16}$
\par}
\vspace{0.5cm}
{\footnotesize
\begin{flushright}
\begin{minipage}{0.75\textwidth}
$^{1}$ CERN, Switzerland \\[0.2em]
$^{2}$ Lawrence Berkeley National Laboratory, USA \\[0.2em]
$^{3}$ Institute of High Energy Physics, China \\[0.2em]
$^{4}$ Centre de Physique des Particules de Marseille CPPM, France \\[0.2em]
$^{5}$ CNRS/IN2P3 and Aix-Marseille Université, France \\[0.2em]
$^{6}$ University of Manchester, UK \\[0.2em]
$^{7}$ University of Michigan, USA \\[0.2em]
$^{8}$ Brookhaven National Laboratory, USA \\[0.2em]
$^{9}$ University of Notre Dame, USA \\[0.2em]
$^{10}$ Bethel University, USA \\[0.2em]
$^{11}$ Paul Scherrer Institute, Switzerland \\[0.2em]
$^{12}$ Helsinki Institute of Physics, Finland \\[0.2em]
$^{13}$ California Institute of Technology, USA \\[0.2em]
$^{14}$ KEK, Japan \\[0.2em]
$^{15}$ Princeton University, USA \\[0.2em]
$^{16}$ Deutsches Elektronen-Synchrotron DESY, Germany \\[0.2em]
$^{17}$ Fermi National Accelerator Laboratory, USA \\[0.2em]
$^{18}$ Indian Institute of Science and Education (Pune), India \\[0.2em]
$^{19}$ Federal University of Alfenas, Brazil \\[0.2em]
$^{20}$ Humboldt University of Berlin, Germany \\[0.2em]
\end{minipage}
\end{flushright}
}
\vspace{0.5cm}
{\large \today}

{\normalsize Version: v1.0-16a371e}

{\normalsize These recommendations are best read on the web site:\\
\url{https://icfa-data-best-practices.app.cern.ch/}}
\vspace{1cm}
\vfill
\vspace{-2cm}
\noindent\rule{5cm}{0.4pt}
\\
{\footnotesize $^{*}$ Corresponding editor: Kati Lassila-Perini (contact via icfa-data-best-practices-contact at cern.ch)}
\vspace{1cm}
\end{center}
\end{titlepage}
\let\cleardoublepage\clearpage

\chapter*{Foreword}

These recommendations are the result of reflections by scientists and experts who are, or have been, involved in the preservation of high-energy physics data. The work has been done under the umbrella of the Data Lifecycle panel of the International Committee of Future Accelerators (ICFA), drawing on the expertise of a wide range of stakeholders.

A key indicator of success in the data preservation efforts is the long-term usability of the data. Experience shows that achieving this requires providing a rich set of information in various forms, which can only be effectively collected and preserved during the period of active data use.

The recommendations are intended to be actionable by the indicated actors and specific to the particle physics domain. They cover a wide range of actions, many of which are interdependent. These dependencies are indicated within the recommendations and can be used as a road map to guide implementation efforts.

It is useful to note here that, for HEP experiments, data cannot be separated from the entire processing environment, which includes all necessary digital or virtual information associated with scientific information extraction. This processing environment encompasses metadata, software (including analysis environments), databases, and documentation. It also, in an extended sense, covers publication procedures and legal aspects related to property, knowledge transfer, and similar concerns. All these elements must be considered when planning operations that extend beyond the boundaries of a single collaboration, such as long-term data preservation and open data initiatives.

Similarly, “preservation” or “opening” are well-defined and usually highly specialised technological projects, including a significant amount of design and R\&D, that have to be planned and deployed in a professional manner, as is the case with any other computing activity in HEP. They will not just occur spontaneously and neglecting them has consequences, the overarching one being the loss of physics potential.

Some elements of these recommendations refer to the four levels of open data defined by the Data Preservation for High Energy Physics collaboration:

\begin{itemize}
\item \textbf{Level 1:} Data in support of publications (e.g. digitized plots or HepData records)
\item \textbf{Level 2:} Data for education and outreach (e.g. simplified ntuples or event data)
\item \textbf{Level 3:} High-level data for research (experiment analysis formats)
\item \textbf{Level 4:} Raw data from the detector.
\end{itemize}

This document and its recommendations focus on open science and data preservation issues for the purposes of scientific research. Other aspects of open science, including open data for education and outreach, overlap with the issues discussed here but should be evaluated and addressed separately.

The Data Lifecycle panel plans to conduct a follow-up evaluation of the current situation in the field.

\newpage

\tableofcontents
\newpage

\chapter{Executive Summaries}

\section{Executive Summary for Host laboratory}
Host laboratories serve as critical institutional anchors for open science and long-term data preservation in high-energy physics. They provide lasting infrastructure, governance frameworks, and custodial oversight that extend beyond the duration of individual experimental collaborations. To fulfill this stewardship role, host laboratories must work closely with experimental collaborations and be supported by sustained funding from funding agencies.
\subsection*{Key actions}
\subsubsection*{1. Establish clear policies}
Develop comprehensive archival (\hyperref[sec:PM1]{PM1}) and open science (\hyperref[sec:PM2]{PM2}) policies that encompass all relevant research outputs—data, software, and documentation. These policies should be developed in partnership with experimental communities, reviewed regularly to stay current with evolving best practices, and aligned with national and funding agency requirements. Ensure effective communication and proper implementation of policies across collaborations (\hyperref[sec:PM16]{PM16}).
\subsubsection*{2. Ensure long-term institutional continuity}
Coordinate with experimental management to develop succession plans for data stewardship (\hyperref[sec:PM13]{PM13}), web presence (\hyperref[sec:PM14]{PM14}), and infrastructure oversight. Create strategies for preserving collaboration metadata not covered in other plans (\hyperref[sec:PM15]{PM15}) and establish clear procedures for transitioning from active operations to long-term custodianship beyond the experiment’s duration.
\subsubsection*{3. Provide essential infrastructure}
Maintain critical infrastructure for long-term preservation, including: version control systems (\hyperref[sec:IR1]{IR1}), software archival services (\hyperref[sec:IR2]{IR2}), open data repositories (\hyperref[sec:IR3]{IR3}), web hosting and archiving (\hyperref[sec:IR4]{IR4}), and conditions databases and related services (\hyperref[sec:IR5]{IR5}). These systems must support both ongoing activities and long-term access, with service levels designed to extend beyond experiment operations.
\subsubsection*{4. Coordinate sustainable resources}
Work with funding agencies to secure dedicated and sustainable funding for infrastructure, staffing, and operations (\hyperref[sec:CF1]{CF1}). Implement dual-level funding mechanisms for experiment-level and laboratory-level support (\hyperref[sec:CF2]{CF2}). Coordinate storage planning and custodial agreements with collaborations to ensure long-term accessibility of internal and public data (\hyperref[sec:PM13]{PM13}), recognizing that preservation costs continue beyond the active research phase.
\subsubsection*{5. Practice active technology stewardship}
Assign dedicated resources to anticipate and manage infrastructure evolution: Monitor legacy data access risks (\hyperref[sec:LS2]{LS2}), archive external software dependencies where permitted (\hyperref[sec:LS3]{LS3}), manage changes in storage infrastructure (\hyperref[sec:LS4]{LS4}), maintain technological vigilance to anticipate disruptions and guide long-term strategies (\hyperref[sec:LS5]{LS5}), and implement regular monitoring and evaluation of preservation progress across experiments (\hyperref[sec:CF6]{CF6}).
\subsection*{Expected outcomes}
By implementing these coordinated practices, host laboratories will serve as trusted guardians of scientific knowledge. These measures will keep research outputs accessible and reusable, strengthen institutional credibility, and show accountability for public investment.
Host laboratories that lead in open science and long-term preservation will help develop a global research environment rooted in transparency, reusability, and lasting impact. Success depends on ongoing multi-institutional coordination (\hyperref[sec:IC1]{IC1})  and resources throughout and beyond the experimental lifecycle, as outlined in funding and coordination recommendations (\hyperref[sec:CF1]{CF1}, \hyperref[sec:CF2]{CF2}, \hyperref[sec:PM13]{PM13}, \hyperref[sec:PM17]{PM17}).

\section{Executive Summary for Experiment management}
Long-term data preservation is critical to ensure the scientific legacy of high-energy physics (HEP) experiments. Actions supporting data preservation must begin during active data-taking and analysis periods to maintain the usability and value of data and associated knowledge for future generations.
\subsection*{Unprecedented data volumes and scientific potential}
Modern HEP experiments generate unprecedented volumes of data with extraordinary complexity and richness. These datasets represent a substantial investment and contain potential for studies that extend far beyond the original experimental program. The full scientific value of these data can only be realized through careful preservation that allows for reanalysis as theoretical understanding and computational techniques advance over the coming decades.
\subsection*{Planning and resources}
Realizing this potential requires robust, long-term planning for storage, computing infrastructure, and human resources, along with regular reassessment to address evolving data volumes and technologies. Early and clear agreements on the transfer of custodial responsibility—from the experiment to the host laboratory or a national data archive—must be established and maintained. Continuous open data releases, supported by dedicated expertise and infrastructure, ensures that datasets remain accessible and relevant for future generations of researchers.
\subsection*{Preserving analysis knowledge}
While the HEP community has set strong standards for open access publishing and made progress in public data releases, there remain substantial challenges in preserving analysis knowledge—the workflows, software, and contextual documentation required for true data reusability.
To address these challenges, these recommendations set as a goal the establishment of preserved, FAIR (Findable, Accessible, Interoperable, and Reusable) analysis workflows as standard practice within the experiment. By adopting these guidelines, the analysis process within the collaboration—with access to the input data—becomes FAIR, ensuring analyses can be efficiently located, understood, and reused by collaborators, and laying the groundwork for extending these benefits when data and materials are released more broadly. The recommendations further advocate making publication-related software, analysis workflow descriptions, and environment specifications publicly available at the time of publication, encouraging open understanding of the research process and making future reuse possible.
\subsection*{Key actions}
\subsubsection*{1. Practical preservation planning (PM3, PM5, PM7, PM13, DM4):}
Create and maintain clear plans describing what data, software, and documentation will be preserved and shared; where and how these will be stored; who is responsible; and how custodial responsibility and resource commitments are secured both during and after the collaboration, in cooperation with the host laboratory.
\subsubsection*{2. Enable open sharing  (PM5, PM6, PM7, PM8, PM11,AP11):}
Make regular, high-quality releases of event-level data a core element of the experiment’s open science program. Ensure that publication-related data, software, and analysis workflows are prepared and shared as part of the standard publication process. Provide hands-on support to analysts so all these research products are well documented, properly archived, and reusable.
\subsubsection*{3. Recognize and resource (PM9, PM10, IC1, IC2):}
Formally recognize data preservation and open science activities as essential research tasks. Allocate stable resources and include these responsibilities in official roles and succession plans at both experiment and institutional levels.
\subsubsection*{4. Support skills and best practices (SK1, SK2, SK6, PM9):}
Offer regular training, mentorship, and practical guidance to foster software and data preservation skills. Encourage and incentivize adoption of open science best practices in everyday research.
\subsubsection*{5. Ensure sustainable knowledge infrastructure (PM14, PM15, PM16):}
Plan for the long-term preservation and accessibility of the collaboration’s web resources, metadata, and communication channels. Guarantee that documentation, data, and analysis materials remain discoverable and understandable for future users.
\subsection*{Expected outcomes}
The active encouragement and enforcement of these recommendations by experiment management is essential for their success. When collaboration members follow these best practices in their daily analysis work, knowledge sharing and working efficiency are significantly improved. Researchers develop valuable skills in collaborative software development and data analysis work—skills that are widely applicable both within and beyond the scientific community.
At the same time, these practices help keep data usable and analysis knowledge available over time, allowing future collaboration members and the wider public to use them for new research. The extent of adoption will naturally vary depending on the size and resources of each experiment, but every step in this direction adds lasting value.

\section{Executive Summary for Home institute}
The long-term value of scientific data depends not only on the efforts of experiment collaborations and the host laboratory, but also on the active support and clear expectations of home institutes. By prioritizing effective training, encouraging contributions to community resources, and valuing open science work, home institutes play a direct role in sustaining high research standards and ensuring data and knowledge remain useful over time.
\subsection*{Key actions}
\subsubsection*{1. Support software skills development:}
Allocate time in work plans and integrate training into onboarding, curricula, and professional development to build up-to-date computing skills (\hyperref[sec:SK2]{SK2}, \hyperref[sec:SK3]{SK3}, \hyperref[sec:SK8]{SK8}, \hyperref[sec:SK7]{SK7}).
\subsubsection*{2. Encourage contributing to common tools development:}
Motivate and support researchers to develop and improve shared community tools and participate in community training initiatives (\hyperref[sec:SK4]{SK4}, \hyperref[sec:PM10]{PM10}).
\subsubsection*{3. Establish expectations for preservation of publication-related software:}
Set requirements for preserving and making accessible code, workflows, and documentation linked to publications (\hyperref[sec:PM12]{PM12}).
\subsubsection*{4. Give value to open science activities:}
Recognize and reward software skills training, documentation, and tool development in performance evaluations and academic credits (\hyperref[sec:SK5]{SK5}, \hyperref[sec:DK2]{DK2}, \hyperref[sec:PM10]{PM10}, \hyperref[sec:SK7]{SK7}, \hyperref[sec:IC1]{IC1}).
\subsubsection*{5. Track and demonstrate career impact:}
Monitor and record the career progression of employees who develop software skills to demonstrate the value of these competencies for future opportunities (\hyperref[sec:CF5]{CF5}).
\subsection*{Expected outcomes}
These actions ensure that researchers systematically develop strong software and computing skills—capabilities fundamental for current scientific work and highly valuable for future careers in academia, industry, and technology sectors. By reinforcing these practices, home institutes guarantee that data and analysis knowledge remain accessible and usable to future collaboration members and the broader public for new research long after the original studies conclude.

\section{Executive Summary for WG leaders}
Have you encountered challenges such as:
\begin{itemize}
  \item Last-minute plot changes for publications being blocked because the precise version of code or input configurations was missing?
  \item Crucial knowledge being lost when a key analyst or developer transitions away, leaving code or documentation unclear or incomplete?
  \item Difficulty recombining or reusing past analysis results because necessary code, workflows, or system details were inaccessible?
\end{itemize}
These are not uncommon issues in high-energy physics collaborations, and this set of best practice recommendations directly addresses them. Adopting these practices yields immediate benefits: reducing duplicated effort, preventing delays, and ensuring that knowledge remains available to support both current and future work. While the larger goals of data preservation and open science may appear remote, implementing FAIR practices—making code, workflows, and documentation Findable, Accessible, Interoperable, and Reusable—will make your group’s daily activities more efficient and sustainable.
\subsection*{Your role}
As a working group leader, your influence shapes research culture and daily standards. You may not be an expert in every tool or workflow, and concerns about transition costs or initial disruptions are understandable. However, these recommendations reflect modern, widely accepted software and collaboration practices that ultimately streamline both analysis and development work.
\subsection*{Key actions:}
\subsubsection*{1. Common tools:}
Promote the use of and contribution to common tools and workflows to support collaboration and reduce duplication (\hyperref[sec:AP3]{AP3}, \hyperref[sec:AP5]{AP5}, \hyperref[sec:AP6]{AP6}, \hyperref[sec:DK6]{DK6}).
\subsubsection*{2. Reproducibility standards:}
Set clear expectations for documentation and reproducibility from the start of every project, and make sure code and workflow definitions are properly versioned, archived and accessible (\hyperref[sec:AP1]{AP1}, \hyperref[sec:AP2]{AP2}, \hyperref[sec:AP4]{AP4}, \hyperref[sec:AP7]{AP7}, \hyperref[sec:AP8]{AP8}, \hyperref[sec:AP12]{AP12}, \hyperref[sec:AP14]{AP14}).
\subsubsection*{3.Open culture:}
Foster a culture of openness, collaboration, and training, including lowering barriers for new contributors and supporting ongoing skills development (\hyperref[sec:SK1]{SK1}, \hyperref[sec:SK4]{SK4}, \hyperref[sec:CS2]{CS2}).
\subsubsection*{4. Advocate FAIRness}
Actively advocate for and support FAIR practices within your group and the wider collaboration (\hyperref[sec:PM9]{PM9}, \hyperref[sec:AP11]{AP11}, \hyperref[sec:IC2]{IC2}).
Your leadership is essential for turning these recommendations into everyday habits, ensuring your group’s work remains valuable and usable—both now and for the future.

\section{Executive Summary for Funding agency}
Funding agencies should implement comprehensive policies to guarantee the long-term accessibility, usability, and impact of publicly funded particle physics research. Allocating funding for open science and sustained data preservation is essential to ensure that valuable research data and software remain accessible, reusable, and verifiable over the long term. Dedicated resources enable the development and maintenance of robust infrastructures, support compliance with best practices, and maximize the scientific and societal return on investment by facilitating future discoveries, transparency, and collaboration. The following actions are crucial for maximizing scientific returns and promoting open science practices.
\subsection*{Key actions}
\subsubsection*{1. Planning requirements [PM17, DK8, AP15]}
Require all experiment funding applications to include comprehensive data management and open science plans. These should cover FAIR principles, detail preservation strategies, define responsibilities, and allocate sufficient budgets for both host laboratory infrastructure and experiment-specific requirements.
\subsubsection*{2. Software management and training [SK6, SW12]}
Require software training plans that emphasize data management best practices, and software management plans that ensure proper code documentation, sharing, and long-term preservation. Funded projects must treat research software as a core scientific product. Funding agency guidelines can play a crucial role by setting clear expectations and standards.
\subsubsection*{3. Dual-level funding mechanisms [PM17, CF1, CF2, DK8]}
Provide coordinated funding at both the experiment and laboratory levels—supporting active data preservation during experiments and ensuring long-term infrastructure maintenance after projects end. The relevant costs are small compared to the original investment in the infrastructure represented by the construction of an experiment and the operation resulting in data collection, and will result in the long-term availability of the data for future use.
\subsubsection*{4. Broaden evaluation criteria [PM18]}
Adopt review criteria that value contributions to open science, data preservation, software development, training, and outreach alongside traditional research metrics. This approach promotes practices essential for reproducible and transparent science.
\subsection*{Expected outcomes}
These policies will ensure that decades of research results remain accessible for future discoveries. They will promote cultural change toward sustainable, open research practices while demonstrating accountability for public funds. They also improve transparency, reproducibility, and the societal value of particle physics research.
By establishing these standards, funding agencies can guide the scientific community toward more effective and sustainable research practices that protect the long-term value of public research investments.

\section{Executive Summary for Tool developers}
The scientific community depends on the expertise and dedication of tool developers to create and maintain both community-wide and experiment-specific tools with different scopes and user communities. These software tools are essential for everyday work, making data analysis procedures more efficient and helping to avoid duplication of effort. The long-term usability of scientific data depends directly on the continued availability or proper preservation of these tools. Many leading tool developer groups are already models of best practices in open science, setting high standards that benefit the broader community.
\subsection*{Key actions}
\subsubsection*{1. Release open-source software and apply an OSI-approved license:}
Make all tools openly available with clear, OSI-approved licensing, ensuring legal clarity and fostering reuse (\hyperref[sec:LC2]{LC2}, \hyperref[sec:LC4]{LC4}, \hyperref[sec:LC5]{LC5}).
\subsubsection*{2. Provide clear copyright statements:}
Include accurate copyright information in code and documentation in line with host laboratory and home institute guidelines (\hyperref[sec:LC1]{LC1}).
\subsubsection*{3. Document how to contribute:}
Offer clear contribution guidelines and templates in all repositories to make it easier for new contributors to get involved (\hyperref[sec:CS2]{CS2}).
\subsubsection*{4. Ensure long-term availability:}
Archive all source code, maintain comprehensive documentation for each release, and support legacy formats and data migration when relevant (\hyperref[sec:CM1]{CM1}, \hyperref[sec:CM3]{CM3}, \hyperref[sec:CM4]{CM4}, \hyperref[sec:CS3]{CS3}, \hyperref[sec:CS4]{CS4}, \hyperref[sec:CS5]{CS5}).
\subsubsection*{5.Provide container images:}
Package software in standard OCI-compliant containers to ensure consistent, reproducible environments across different computing platforms (\hyperref[sec:CM2]{CM2}).
\subsection*{Expected outcomes}
By following these practices, tool developers safeguard the accessibility and usability of both the software and the data it supports for years to come. This maximizes the value of their work for current and future researchers—both within collaborations and in the wider scientific community—by making  critical research tools easy to use, adapt, and build upon.

\section{Executive Summary for Analysts}
\begin{itemize}
  \item Have you ever forgotten where you put your code? (\hyperref[sec:AP1]{AP1}, \hyperref[sec:SW1]{SW1}, \hyperref[sec:SW2]{SW2})
  \item Ever wondered what the exact version was that produced a set of plots? (\hyperref[sec:AP7]{AP7}, \hyperref[sec:SW11]{SW11})
  \item Have you ever looked at code a colleague shared but couldn’t figure out where or how to run it? (\hyperref[sec:AP12]{AP12}, \hyperref[sec:DK1]{DK1}, \hyperref[sec:SW8]{SW8})
  \item Have you ever broken your functioning code with some updates and spent days fixing it? (\hyperref[sec:SW9]{SW9}, \hyperref[sec:SW3]{SW3}, \hyperref[sec:AP3]{AP3})
  \item Have you ever written a piece of code, only to discover that a colleague had already done it, and using theirs would have saved you days? (\hyperref[sec:AP5]{AP5}, \hyperref[sec:CS2]{CS2})
  \item Or have you ever found code you could have used, but adapting it seemed too much trouble? (\hyperref[sec:AP5]{AP5}, \hyperref[sec:SW6]{SW6}, \hyperref[sec:DK1]{DK1})
  \item Have you ever tried to run your code after a few months’ break, only to find that your environment had changed and the code no longer runs? (\hyperref[sec:AP12]{AP12}, \hyperref[sec:SW8]{SW8})
  \item Have you ever spent hours typing commands and running scripts one after another thinking that it really could have been automated? (\hyperref[sec:AP6]{AP6}, \hyperref[sec:SW10]{SW10})
\end{itemize}
\subsection*{Yes?}
These recommendations are for you. Data preservation and open science may sound like distant goals with little relevance to your daily work. But following these recommendations brings immediate benefits: for you, for your working group, and for your experiment. These recommendations emphasize that sharing code and workflows should become standard practice in our community, and that experiments should make sharing code and workflows with publications an expectation, not an exception.
\subsection*{What is it not?}
Worried that someone will inspect your code and judge your coding? That’s not the goal. The goal is for your code to serve as a precise and unambiguous description of your analysis procedure. Concerned you’ll need to spend weeks building a user manual and documentation? No - you only need to document what your code does and how to run it, well enough for, for example, a newcomer in your group to start using it. Do you need to commit to maintaining your code in the future? Not unless you want to.
\subsection*{Key actions}
\subsubsection*{1. Develop your software skills:}
Make use of version control, reproducible environments, and best practices for documentation and automation (\hyperref[sec:AP2]{AP2}, \hyperref[sec:SW2]{SW2}, \hyperref[sec:SW6]{SW6}, \hyperref[sec:SW9]{SW9}, \hyperref[sec:AP12]{AP12}, \hyperref[sec:SW8]{SW8}).
\subsubsection*{2.Adapt your working habits so that preserving your analysis work becomes routine:}
Use common infrastructure and templates for configuration, and testing. Regularly document your software environment and ensure your code and workflows are easy to understand and run by others (\hyperref[sec:AP1]{AP1}, \hyperref[sec:AP3]{AP3}, \hyperref[sec:AP4]{AP4}, \hyperref[sec:AP7]{AP7}, \hyperref[sec:AP9]{AP9}, \hyperref[sec:AP12]{AP12}, \hyperref[sec:AP14]{AP14}, \hyperref[sec:DK1]{DK1}, \hyperref[sec:LC2]{LC2}, \hyperref[sec:LC5]{LC5}).
\subsubsection*{3. Contribute to common tools and share improvements:}
Prefer using, extending, and contributing to collaboration-wide analysis tools and software, and help improve shared documentation (\hyperref[sec:AP5]{AP5}, \hyperref[sec:CS2]{CS2}, \hyperref[sec:SW6]{SW6}).
\subsection*{Your role}
Achieving the goal of making your results FAIR (Findable, Accessible, Interoperable, and Reusable)—first of all, for future you, and for your working group, then for your experiment, and for the scientific community — is the result of many small and interconnected actions. As analysts, your margin of maneuver often depends on the decisions and support provided by your home institute and experiment management—they can give you the right conditions and resources to make this possible. However, when it comes to open science, and especially to sharing your code, no one else can do it for you. The final step is always yours: only you can ensure your code and analysis practices are open, understandable, and reusable by others.

\section{Executive Summary for Data management}
The availability and usability of experimental data rely on the expertise and commitment of the data management team, whose daily work ensures that data is organized, stored, preserved, and made accessible for both current analyses and future research. While the main focus is often on everyday and short- to medium-term operations, the team’s actions have a significant impact on data reusability throughout its entire lifecycle.
Experience shows that every decision made today—how datasets are processed, catalogued, accompanied by metadata, and linked with supplementary information—directly shapes the possibilities for tomorrow’s science. Keeping these recommendations in mind during design choices and key decisions helps data management teams ensure the availability and interpretability of results for years to come.
\subsection*{Key actions}
\subsubsection*{1. Define and document legacy datasets:}
For each data-taking period, establish a clear, comprehensive definition of the legacy datasets—including collision data and simulations—with detailed documentation of how they can be identified, accessed, and analyzed (\hyperref[sec:DM1]{DM1}).
\subsubsection*{2. Safeguard supplementary and contextual data:}
Preserve all supporting information essential for data interpretation and reuse, such as processing workflows, event selections, luminosity details, calibration, quality indicators, and file catalogs (\hyperref[sec:DM2]{DM2}, \hyperref[sec:DM3]{DM3}).
\subsubsection*{3. Record data processing details:}
Systematically archive information on every processing step, including trigger and generator configurations, software versions, input parameters, and any runtime conditions for both real and simulated data (\hyperref[sec:DM5]{DM5}).
\subsubsection*{4. Ensure rich, accessible metadata and software availability:}
Associate each preserved dataset with comprehensive, machine-readable metadata, and provide documented access to all necessary software and instructions for data analysis, adapted to evolving research contexts (\hyperref[sec:DM7]{DM7}).
\subsubsection*{5. Guarantee recoverability and reprocessing:}
Maintain and regularly test the full data processing and analysis environment—including all software, workflows, and required supplementary data—to guarantee that legacy datasets can be regenerated or restored in case of data loss (\hyperref[sec:DM9]{DM9}).
\subsubsection*{6. Favor community-standard formats:}
Whenever possible, structure and deliver data so it can be accessed and analyzed using widely adopted community tools, minimizing dependence on experiment-specific software and enhancing future interoperability (\hyperref[sec:DM11]{DM11}).
\subsection*{Expected outcomes}
By following these principles and recommendations, data management teams ensure that high-quality data and its associated knowledge remain useful and accessible—maximizing the return on investment for the entire scientific community. Their work empowers both current collaborations and future generations of researchers to unlock the full scientific potential of experimental data, ensuring that data collected today remain understandable and reusable for decades to come.

\section{Executive Summary for Open data group}
The open data group plays a central role in implementing the experiment’s open data policy, enabling access to experimental datasets for the wider scientific community and the public. By acting as a vital bridge between the experiment, the host laboratory’s services, and the community of open data users, the group ensures not only the availability of data but also its meaningful and effective use far beyond the original collaboration.
While much of the group’s daily work focuses on the technical and procedural challenges of preparing and releasing high-quality data, their actions have a significant and lasting influence on the discoverability, transparency, and reusability of the experiment’s data. Thoughtful choices about formats, metadata, supporting software, and user engagement today directly determine how easily the data can be discovered, understood, and reused by diverse audiences in the future.
The successful work of the open data group, however, depends on the broad adoption of best practices by all members of the collaboration as well as the availability of sufficient resources to support and maintain preserved data. Strong support from experiment management and laboratory services is equally essential to ensure that open data releases remain reliable, useful, and sustainable over time.
\subsection*{Key actions}
\subsubsection*{1. Publish open data in recognized repositories:}
Release event-level experimental data and simulations in repositories that align with the experiment’s open science policy and institutional context, ensuring open, authentication-free access (\hyperref[sec:DM6]{DM6}).
\subsubsection*{2. Provide rich metadata and supporting software:}
Accompany open data with thorough, machine-readable metadata—including content, processing, and contextual information—and make all necessary analysis software and usage guidelines openly available and well documented (\hyperref[sec:DM7]{DM7}).
\subsubsection*{3. Use standardized formats and persistent identifiers:}
Distribute open data in widely adopted, standardized formats and ensure assignment of persistent identifiers (DOIs), for discoverability, interoperability, and long-term access (\hyperref[sec:DM10]{DM10}).
\subsubsection*{4. Offer example analysis workflows:}
Supply open datasets with introductory and publication-linked analysis examples, designed to help both general and expert users quickly understand and make use of the data (\hyperref[sec:DK7]{DK7}).
\subsubsection*{5. Engage in user-focused reproducibility campaigns:}
Organize regular campaigns in which external scientists test the capability to reproduce analyses using open data, incorporating their feedback to improve usability and documentation (\hyperref[sec:DM8]{DM8}).
\subsection*{Expected outcomes}
Through these practices—with the support and engagement of the entire collaboration and experiment management—the open data group ensures accessibility and reusability of the experiment’s data. By fostering productive exchanges between the experiment, laboratory services, and the wider user community, they help maximize the long-term value and reach of the data, supporting their current and future use.

\newpage

\chapter{PM: Policy and management}

\section{PM1: Develop a comprehensive archival policy that encompasses all relevant scientific research outputs generated through the organization’s research activities, supported by clear governance frameworks and technical safeguards to ensure their long-term preservation.}
\label{sec:PM1}

\subsection*{Description}
\begin{itemize}
  \item Description: A robust archival policy is essential for preserving the scientific legacy of research conducted at the laboratory. It should explicitly include all major outputs, publications, research methods (including software, computing environments, and workflows), datasets, technical documentation of experiments, and engineering records. The policy must specify ownership, custodial responsibilities, and inheritance procedures, especially in collaborative settings involving multiple institutions. Clear communication channels between the laboratory and experimental collaborations are essential to ensure coordinated preservation and prevent gaps in coverage. Sustained accessibility relies on dependable, redundant storage systems and regularly tested disaster recovery plans. Without a comprehensive policy, valuable research assets risk being lost due to unclear responsibilities, infrastructure shortcomings, or lack of institutional commitment. As the foundation for all open science and preservation efforts, the archival policy offers the structure and accountability needed for the long-term, responsible management of scientific knowledge.
\end{itemize}

\subsection*{Class}
\begin{itemize}
  \item Policy and management
\end{itemize}

\subsection*{Actors}
\begin{itemize}
  \item host laboratory
\end{itemize}

\subsection*{Enables}
\begin{itemize}
  \item \hyperref[sec:PM13]{PM13}
  \item \hyperref[sec:PM9]{PM9}
  \item \hyperref[sec:PM3]{PM3}
  \item \hyperref[sec:PM17]{PM17}
  \item \hyperref[sec:PM16]{PM16}
  \item \hyperref[sec:PM14]{PM14}
  \item \hyperref[sec:PM19]{PM19}
  \item \hyperref[sec:DK4]{DK4}
\end{itemize}

\newpage

\section{PM2: If not directly governed by national open science policies, establish a comprehensive open science policy that commits the laboratory to making all relevant research outputs, including datasets, related software, and research findings, publicly accessible and freely available.}
\label{sec:PM2}

\subsection*{Description}
\begin{itemize}
  \item Description: A comprehensive open science policy affirms the laboratory’s commitment to transparency, accessibility, and the long-term value of research conducted at its facilities. Developed collaboratively with experiments and stakeholders, it should include open access to publications, data, and software; education and outreach; research integrity; and the infrastructure and career policies necessary to support these practices.
  \item Description: The policy must adhere to FAIR principles and include an implementation plan with clear resources, timelines, and communication channels. Regular progress reviews ensure ongoing relevance and effectiveness. Without such planning and engagement, policies risk becoming symbolic rather than actionable.
\end{itemize}

\subsection*{Class}
\begin{itemize}
  \item Policy and management
\end{itemize}

\subsection*{Actors}
\begin{itemize}
  \item host laboratory
\end{itemize}

\subsection*{Enables}
\begin{itemize}
  \item \hyperref[sec:CF6]{CF6}
  \item \hyperref[sec:PM5]{PM5}
  \item \hyperref[sec:PM13]{PM13}
  \item \hyperref[sec:PM9]{PM9}
  \item \hyperref[sec:PM3]{PM3}
  \item \hyperref[sec:PM17]{PM17}
  \item \hyperref[sec:CF2]{CF2}
  \item \hyperref[sec:PM16]{PM16}
  \item \hyperref[sec:PM11]{PM11}
  \item \hyperref[sec:PM7]{PM7}
  \item \hyperref[sec:CF1]{CF1}
\end{itemize}

\newpage

\section{PM3: Agree on and approve a data management plan that covers long-term data preservation.}
\label{sec:PM3}

\subsection*{Description}
\begin{itemize}
  \item Description: An agreed-upon and officially approved data management plan sets the stage for responsible long-term handling of research outputs. This plan should cover how legacy data is preserved and kept reusable for future analysis within the experiment and the custodial responsibility of the data and related software for the time after the collaboration ceases to exist. This should ideally be taken by the original host site of the experiment, or, in the case of a national laboratory, by a national-level data archiving infrastructure, and needs to be discussed and agreed with these stakeholders in agreement with the host laboratory’s archival policy. The plan should also describe the process for transitioning custodial responsibility and resource commitments beyond the lifetime of the experiment. It should recognize that formal agreements for the post-experiment period may need to be developed later, as clearer information about future data volumes and preservation costs becomes available.
  \item Description: The data management plan should address not only the raw experimental data and its preservation but also the preservation and publication of analysis-level or higher-level data, along with essential metadata (e.g., conditions data, calibration data, publication metadata, dataset metadata). It should also ensure the ability to recreate analysis-level data formats from raw data and to generate simulated data, for example, in the event of accidental data loss.
  \item Description: The conservation of supporting documentation, including such things as historical records, configuration descriptions, analysis recommendations, users’ manuals, and any other elements required to understand and re-use the preserved data should also be considered in the data management plan. Consideration should be given to long-term authorization issues: who should have permission to access formerly-internal documents after the collaboration has concluded.
  \item Description: The long-term data preservation aspects in the data management plan should be reviewed regularly with respect to implementation status, regularly updated estimates of data volume, and technological evolution. Experiments should maintain and refine their estimates of the total amount of data to be preserved—including raw data and analysis data formats including simulations—throughout the lifetime of the experiment.
  \item Description: Open data (\hyperref[sec:PM5]{PM5}) and software management (\hyperref[sec:PM7]{PM7}) policies can be part of this plan or defined as separate documents. While open data policies may already be in place for ongoing experiments, software as a part of open research products is an emerging practice and a roadmap of actions for enabling it is provided in these recommendations.
\end{itemize}

\subsection*{Class}
\begin{itemize}
  \item Policy and management
\end{itemize}

\subsection*{Actors}
\begin{itemize}
  \item experiment management
\end{itemize}

\subsection*{Dependencies}
\begin{itemize}
  \item \hyperref[sec:PM2]{PM2}
  \item \hyperref[sec:PM1]{PM1}
\end{itemize}

\subsection*{Enables}
\begin{itemize}
  \item \hyperref[sec:AP3]{AP3}
  \item \hyperref[sec:PM19]{PM19}
  \item \hyperref[sec:PM16]{PM16}
  \item \hyperref[sec:PM17]{PM17}
\end{itemize}

\newpage

\section{PM4: Mandate that the data management team of the experiment be explicitly responsible for overseeing long-term preservation aspects.}
\label{sec:PM4}

\subsection*{Description}
\begin{itemize}
  \item Motivation: Experiments’ data management teams usually cover the operational aspects, but they are often focused on current, short- and medium-term needs, whereas long-term data preservation requires dedicated attention and planning that extends beyond the active lifetime of the collaboration. Experiment management should address this by explicitly including long-term preservation in the data management team’s mandate, and by following up on progress and allocating appropriate resources.
\end{itemize}

\subsection*{Class}
\begin{itemize}
  \item Policy and management
\end{itemize}

\subsection*{Actors}
\begin{itemize}
  \item experiment management
\end{itemize}

\subsection*{Enables}
\begin{itemize}
  \item \hyperref[sec:PM6]{PM6}
\end{itemize}

\newpage

\section{PM5: Agree on and approve an open data policy in accordance with the host laboratory’s policy.}
\label{sec:PM5}

\subsection*{Description}
\begin{itemize}
  \item Description: Open data policy is an official commitment by the experimental collaboration, and it must be discussed and agreed on in the governing bodies of the collaboration. It is a long-standing document, and updates should be discussed and decided by the same body that approved the policy.
  \item Description: The policy should cover research-quality event-level data (i.e Level 3 data) and analysis-level data (i.e. Level 1 data such as HEPdata).
  \item Description: The experiment should regularly evaluate the need for and utility of smaller, custom datasets for the broader community.
  \item Description: The experiment policies should be aligned with the host laboratory policy.
  \item Description: The plan should address
  \item What?        Data types, volume, formats.
  \item When?        The schedule for data archiving and sharing.
  \item Where? The intended repositories for archived data and any agreements for their support beyond the lifetime of the experiment.
  \item How?        How the plan enables long- term preservation of the data.
  \item Who?        Roles and responsibilities of the team members in implementing the policy.
  \item Description: Large experiments operate for long times and it is common for formats to evolve based on usage experience and increasing dataset sizes and data volume. Therefore, the details in the open data distribution may evolve.
\end{itemize}

\subsection*{Class}
\begin{itemize}
  \item Policy and management
\end{itemize}

\subsection*{Actors}
\begin{itemize}
  \item experiment management
\end{itemize}

\subsection*{Dependencies}
\begin{itemize}
  \item \hyperref[sec:IR3]{IR3}
  \item \hyperref[sec:PM2]{PM2}
\end{itemize}

\subsection*{Enables}
\begin{itemize}
  \item \hyperref[sec:PM6]{PM6}
  \item \hyperref[sec:DM6]{DM6}
  \item \hyperref[sec:PM16]{PM16}
\end{itemize}

\newpage

\section{PM6: Mandate an open data group to be formed within the experiment to implement the open data policy.}
\label{sec:PM6}

\subsection*{Description}
\begin{itemize}
  \item Motivation: Preparing event-level data releases requires substantial effort. This includes defining the data to be released, collecting metadata and supplementary data products, providing and testing the computing environment setup, and preparing comprehensive documentation. All of these elements must then be organized and formatted appropriately for presentation and access through the chosen data repository. This process requires expertise and a long-term commitment within the experiment.
  \item Motivation: To support the release of analysis-level data, such as HEPdata uploads, it is important that these efforts are actively promoted and sustained by the physics analysis coordination. A dedicated team should be mandated to develop and maintain the necessary tools and procedures, making it easier for analysts to prepare and share their data. This will help ensure that analysis-level data releases become an integrated and efficient part of the publication workflow.
\end{itemize}

\subsection*{Class}
\begin{itemize}
  \item Policy and management
\end{itemize}

\subsection*{Actors}
\begin{itemize}
  \item experiment management
\end{itemize}

\subsection*{Dependencies}
\begin{itemize}
  \item \hyperref[sec:PM4]{PM4}
  \item \hyperref[sec:PM5]{PM5}
\end{itemize}

\subsection*{Enables}
\begin{itemize}
  \item \hyperref[sec:DM6]{DM6}
\end{itemize}

\newpage

\section{PM7: Agree on and approve a software management plan for data and simulation processing software, and for analysis and publication-related code, with the ultimate goal of making all research products—including analysis code—open, and establishing reusable workflows as the standard for all analyses.}
\label{sec:PM7}

\subsection*{Description}
\begin{itemize}
  \item Motivation: Establishing this goal recognizes that software is an essential component of research products, alongside data and results, and that science cannot be fully open without access to code. Making software openly available as a research product is an emerging practice, and these recommendations offer a roadmap of actions for achieving this.
  \item Motivation: Adopting the practice of making code accessible to others brings concrete benefits. In the short term, it streamlines analysis work by making code sharing and collaborative contributions easier within the analysis team. In the longer term, it ensures that the knowledge and technical implementation of the analysis remain accessible for future work within the collaboration. Sharing the code provides a precise and unambiguous description of the analysis procedure, and makes it reusable within the experiment for future analyses and collaborative efforts.
  \item Description: The plan can be part of the Open science and data management plan (\hyperref[sec:PM3]{PM3}) or provided as a separate document, and should contain:
  \item What: Description of management, preservation, and release of software.
  \item When: The schedule for software archiving and sharing.
  \item Where: Location where software will be shared and archived over the long-term.
  \item How: Enable reuse of software through assigning a DOI, license, contribution guidelines, etc.
  \item Who: Roles and responsibilities of the team members.
\end{itemize}

\subsection*{Class}
\begin{itemize}
  \item Policy and management
\end{itemize}

\subsection*{Actors}
\begin{itemize}
  \item experiment management
\end{itemize}

\subsection*{Dependencies}
\begin{itemize}
  \item \hyperref[sec:IR2]{IR2}
  \item \hyperref[sec:PM2]{PM2}
\end{itemize}

\subsection*{Enables}
\begin{itemize}
  \item \hyperref[sec:AP11]{AP11}
  \item \hyperref[sec:PM8]{PM8}
  \item \hyperref[sec:PM16]{PM16}
  \item \hyperref[sec:SK6]{SK6}
  \item \hyperref[sec:AP14]{AP14}
\end{itemize}

\newpage

\section{PM8: Mandate a dedicated group to follow up on the implementation of the software management plan.}
\label{sec:PM8}

\subsection*{Description}
\begin{itemize}
  \item Motivation: Most analysis work is carried out by physicists who are not necessarily experts in software development. To ensure that analysis code can be understood by others and eventually be reusable, researchers need early guidance, practical advice, and access to appropriate tools.
  \item Motivation: Mandating a dedicated group ensures that analysts have access to practical advice and resources for software development from the start. This group can provide templates, tools, and hands-on assistance tailored to the needs of the experiment, making it easier to produce well-structured, shareable code.
\end{itemize}

\subsection*{Class}
\begin{itemize}
  \item Policy and management
\end{itemize}

\subsection*{Actors}
\begin{itemize}
  \item experiment management
\end{itemize}

\subsection*{Dependencies}
\begin{itemize}
  \item \hyperref[sec:PM7]{PM7}
\end{itemize}

\subsection*{Enables}
\begin{itemize}
  \item \hyperref[sec:AP2]{AP2}
  \item \hyperref[sec:AP5]{AP5}
  \item \hyperref[sec:AP3]{AP3}
  \item \hyperref[sec:AP6]{AP6}
\end{itemize}

\newpage

\section{PM9: Act as the advocate, the driver and the sponsor for adopting these recommendations.}
\label{sec:PM9}

\subsection*{Description}
\begin{itemize}
  \item Motivation: These recommendations target different audiences and are designed to be actionable by those groups. Leaders at all levels have a crucial role in making it possible for individuals to take action on these recommendations. To achieve the goals of data preservation and open science, leaders should strive to remove obstacles and actively support implementation. In particular, they should foster a culture that facilitates, rather than hinders, progress toward these goals.
\end{itemize}

\subsection*{Class}
\begin{itemize}
  \item Policy and management
\end{itemize}

\subsection*{Actors}
\begin{itemize}
  \item experiment management
  \item home institute
  \item host laboratory
  \item WG leaders
\end{itemize}

\subsection*{Dependencies}
\begin{itemize}
  \item \hyperref[sec:AP11]{AP11}
  \item \hyperref[sec:AP5]{AP5}
  \item \hyperref[sec:AP8]{AP8}
  \item \hyperref[sec:PM2]{PM2}
  \item \hyperref[sec:PM10]{PM10}
  \item \hyperref[sec:SK2]{SK2}
  \item \hyperref[sec:AP6]{AP6}
  \item \hyperref[sec:DK2]{DK2}
  \item \hyperref[sec:CF1]{CF1}
  \item \hyperref[sec:AP2]{AP2}
  \item \hyperref[sec:SK3]{SK3}
  \item \hyperref[sec:PM12]{PM12}
  \item \hyperref[sec:SK5]{SK5}
  \item \hyperref[sec:SK4]{SK4}
  \item \hyperref[sec:PM1]{PM1}
\end{itemize}

\newpage

\section{PM10: Formally recognize and support the development and maintenance of open science infrastructure and practices as essential research functions.}
\label{sec:PM10}

\subsection*{Description}
\begin{itemize}
  \item Motivation: Open science efforts are often driven by dedicated individuals, even when not formally mandated. Research institutions and experimental collaborations should move beyond relying on individual champions to build sustainable organizational systems for open science:
  \item Description: Home institutes should integrate tasks related to development and maintenance of open science infrastructure and practices into the official work plans of employees.
  \item Description: Experiments should include them in the management structure, and implement succession planning for essential roles to ensure continuity. They should allocate dedicated resources for infrastructure development, maintenance, and evolution as core experimental functions rather than ancillary activities.
  \item Description: Home institutes and the host laboratory should consider joining international efforts towards advancing qualitative and diverse research assessment practices that consider open science activities (e.g., \url{https://coara.eu/}).
\end{itemize}

\subsection*{Class}
\begin{itemize}
  \item Policy and management
\end{itemize}

\subsection*{Actors}
\begin{itemize}
  \item experiment management
  \item home institute
  \item host laboratory
\end{itemize}

\subsection*{Enables}
\begin{itemize}
  \item \hyperref[sec:PM9]{PM9}
\end{itemize}

\newpage

\section{PM11: Publish research results in open-access journals and/or suitable repositories, in accordance with the host laboratory’s policy. Ensure the use of persistent identifiers, comprehensive metadata, appropriate licenses, and proper references to all software and data that the publication relies on.}
\label{sec:PM11}

\subsection*{Description}
\begin{itemize}
  \item Description: The open access version of a paper refers to either the accepted manuscript or the final published article, depending on the host laboratory's policy. It is important that the article includes a software and data availability statement, along with proper references—following standard guidelines (including DOIs)—to the datasets (e.g., Level 1) and software on which it relies.
\end{itemize}

\subsection*{Class}
\begin{itemize}
  \item Policy and management
\end{itemize}

\subsection*{Actors}
\begin{itemize}
  \item experiment management
\end{itemize}

\subsection*{Dependencies}
\begin{itemize}
  \item \hyperref[sec:PM2]{PM2}
\end{itemize}

\newpage

\section{PM12: Ensure employees understand both national and institutional open science requirements, and how compliance can be achieved within international collaborative frameworks.}
\label{sec:PM12}

\subsection*{Description}
\begin{itemize}
  \item Motivation: National open science guidelines often refer to specific national repositories or tools for sharing data and code, which may not be used within large international collaborations. By ensuring that employees are familiar with both national and institutional guidelines, and by following the collaboration’s recommendations, researchers will generally meet national requirements as well. However, it is important for employees to be aware of any additional or more stringent national requirements that may apply, so these can be addressed as needed.
\end{itemize}

\subsection*{Class}
\begin{itemize}
  \item Policy and management
\end{itemize}

\subsection*{Actors}
\begin{itemize}
  \item home institute
\end{itemize}

\subsection*{Enables}
\begin{itemize}
  \item \hyperref[sec:PM9]{PM9}
\end{itemize}

\newpage

\section{PM13: Provide dedicated resources for the long-term preservation of public datasets and coordinate with experiments to maintain storage for internal datasets.}
\label{sec:PM13}

\subsection*{Description}
\begin{itemize}
  \item Motivation: Long-term storage of preserved data is essential for future research. Non-public archived data requires dedicated storage resources, and to ensure data accessibility beyond the experiments’ lifetime, custodial agreements governing long-term preservation need to be established separately between the host laboratory and the experiments. For open data, it is important to secure adequate and separate resources—provided, for example, by the host laboratory—as releases typically occur during the experiments’ active data-taking phase. This guarantees that the usability of open data can be verified while the necessary knowledge is still available, and supports a gradual increase of data volume in line with the host laboratory’s open science and the experiments’ open data policies.
  \item Description: The host laboratory—or, in the case of national laboratories, national-level data infrastructures and services—should agree with the experiments on how existing storage will be maintained beyond the lifetime of the experiment. These coordination efforts should result in agreements between the experiments and the host laboratory—or national infrastructures—that cover the long-term maintenance of storage resources. This is of particular importance in the case of federated storage, which may be located outside the host laboratory, and for which custodial agreements have been made for the active data-taking and analysis phase. In such situations, the host laboratory would typically assume responsibility for long-term preservation unless a common agreement covering federated data preservation resources is in place or being developed.
\end{itemize}

\subsection*{Class}
\begin{itemize}
  \item Policy and management
\end{itemize}

\subsection*{Actors}
\begin{itemize}
  \item host laboratory
\end{itemize}

\subsection*{Dependencies}
\begin{itemize}
  \item \hyperref[sec:CF2]{CF2}
  \item \hyperref[sec:LS4]{LS4}
  \item \hyperref[sec:PM2]{PM2}
  \item \hyperref[sec:PM1]{PM1}
\end{itemize}

\subsection*{Enables}
\begin{itemize}
  \item \hyperref[sec:IR3]{IR3}
  \item \hyperref[sec:DM4]{DM4}
\end{itemize}

\newpage

\section{PM14: Develop and approve a plan for the evolution and migration of collaboration’s information records, such as those for  web presence, after the lifetime of the collaboration, with a focus on long-term viability.}
\label{sec:PM14}

\subsection*{Description}
\begin{itemize}
  \item Motivation: Collaboration information records, such as websites, often become the primary source for understanding legacy data and analyses. Without proper identification and planning, valuable documentation can disappear when hosting arrangements end, making preserved data effectively unusable. Examples from past experiments show that even basic information like detector descriptions and analysis procedures can become inaccessible within 5-10 years without proper archival planning. Consideration should be given to permissions after the collaboration has concluded.
  \item Description: Ideally, this responsibility could be assumed by the original host site of the core web resources, in accordance with the organization's archival policy.
\end{itemize}

\subsection*{Class}
\begin{itemize}
  \item Policy and management
\end{itemize}

\subsection*{Actors}
\begin{itemize}
  \item experiment management
  \item host laboratory
\end{itemize}

\subsection*{Dependencies}
\begin{itemize}
  \item \hyperref[sec:IR4]{IR4}
  \item \hyperref[sec:PM1]{PM1}
\end{itemize}

\subsection*{Enables}
\begin{itemize}
  \item \hyperref[sec:IR4]{IR4}
  \item \hyperref[sec:DK4]{DK4}
\end{itemize}

\newpage

\section{PM15: Develop and approve a comprehensive metadata preservation plan that captures all collaboration information not covered by the data management (PM3), web presence (PM14), or software management (PM7) plans.}
\label{sec:PM15}

\subsection*{Description}
\begin{itemize}
  \item Description: These can be metadata related to specific analysis, such as lead editors or authors, code repositories, and internal documentation. They can include systems through which analysis-critical metadata are provided, e.g. configuration databases, cross-section or calibration data.
\end{itemize}

\subsection*{Class}
\begin{itemize}
  \item Policy and management
\end{itemize}

\subsection*{Actors}
\begin{itemize}
  \item experiment management
  \item host laboratory
\end{itemize}

\subsection*{Dependencies}
\begin{itemize}
  \item \hyperref[sec:AP4]{AP4}
\end{itemize}

\newpage

\section{PM16: Establish and maintain effective communication channels to ensure all collaboration members are informed of and adhere to open science, data preservation, and archival policies.}
\label{sec:PM16}

\subsection*{Description}
\begin{itemize}
  \item Description: Ensure that members of experimental collaborations are familiar with both the experiment’s and the host laboratory’s policies on open science, open data, data preservation, and archiving. Work carried out within collaborations is governed by these rules, which may differ in their practical implementation from the policies of members’ home institutes.
  \item Motivation: Lack of awareness often causes unintentional policy violations that can threaten data preservation, raise legal issues, or hinder open science goals. Posting policies online alone is not enough, collaboration members may miss updates or misunderstand requirements. Clear, frequent, and structured communication helps prevent mistakes and ensures consistent policy implementation across international teams with diverse institutional backgrounds.
\end{itemize}

\subsection*{Class}
\begin{itemize}
  \item Policy and management
\end{itemize}

\subsection*{Actors}
\begin{itemize}
  \item experiment management
  \item host laboratory
\end{itemize}

\subsection*{Dependencies}
\begin{itemize}
  \item \hyperref[sec:PM5]{PM5}
  \item \hyperref[sec:PM2]{PM2}
  \item \hyperref[sec:PM3]{PM3}
  \item \hyperref[sec:PM7]{PM7}
  \item \hyperref[sec:PM1]{PM1}
\end{itemize}

\newpage

\section{PM17: Require comprehensive plans for open science and for data and software management as evaluated components of experiment funding processes, with proper budgeting and clear accountability for both host laboratory infrastructure and experiment-specific needs.}
\label{sec:PM17}

\subsection*{Description}
\begin{itemize}
  \item Description: Data and software management and open science plans should be substantive, reviewed and assessed against best practices and community standards. Effective plans should specify responsibilities for data stewardship, list necessary resources, and detail how FAIR principles will be applied throughout the data lifecycle. They should also cover licensing, data access protocols, strategies for long-term preservation, and coordination between host laboratories and experimental collaborations. Rigorous evaluation ensures proposed approaches are specific, feasible, and aligned with long-term accessibility goals. This process enables funding agencies to prioritize projects with credible preservation strategies and enhances accountability for the long-term value of public research investments.
  \item Description: While laboratory- or experiment-wide data and software management policies should be applied, particular care should be taken for software and data products that might not be considered part of the purview of the experiment, and applicants should be encouraged to exceed the minimum standard set by the laboratory and experiment.
\end{itemize}

\subsection*{Class}
\begin{itemize}
  \item Policy and management
\end{itemize}

\subsection*{Actors}
\begin{itemize}
  \item funding agency
\end{itemize}

\subsection*{Dependencies}
\begin{itemize}
  \item \hyperref[sec:PM3]{PM3}
  \item \hyperref[sec:PM2]{PM2}
  \item \hyperref[sec:PM1]{PM1}
\end{itemize}

\subsection*{Enables}
\begin{itemize}
  \item \hyperref[sec:CF2]{CF2}
\end{itemize}

\newpage

\section{PM18: Create evaluation criteria that explicitly recognize and reward open science contributions alongside traditional research achievements in funding decisions.}
\label{sec:PM18}

\subsection*{Description}
\begin{itemize}
  \item Motivation: Traditional academic evaluations focus on individual publications and first-author results, often overlooking essential work in collaborative infrastructure, data preservation, and community software development. These contributions are vital for sustainable open science but are undervalued in traditional metrics. By formally including open science efforts in proposal reviews, funding agencies can help change research culture and create incentives for researchers to invest in systems and practices that support long-term scientific progress.
\end{itemize}

\subsection*{Class}
\begin{itemize}
  \item Policy and management
\end{itemize}

\subsection*{Actors}
\begin{itemize}
  \item funding agency
\end{itemize}

\newpage

\section{PM19: Plan for the governance of the collaboration after the end of data taking.}
\label{sec:PM19}

\subsection*{Description}
\begin{itemize}
  \item Motivation: A significant portion of an experiment’s scientific output is often realized after data taking ends, provided the data are properly preserved and accessible. However, the end of detector operation represents a disruptive change in the collaboration’s modus operandi, as hardware operation and data collection end. Maintaining a functioning and active governance structure beyond the operational phase is essential for maximizing scientific return. Preparing the process from active data taking to legacy data mode facilitates continued data analysis, supports the publication of new results, and helps attract both interest and funding for further research. Moreover, it ensures that expertise is retained within the collaboration and allows adaptation to new computing and software solutions as technologies evolve, thereby extending the long-term usability of preserved data.
\end{itemize}

\subsection*{Class}
\begin{itemize}
  \item Policy and management
\end{itemize}

\subsection*{Actors}
\begin{itemize}
  \item experiment management
\end{itemize}

\subsection*{Dependencies}
\begin{itemize}
  \item \hyperref[sec:PM3]{PM3}
  \item \hyperref[sec:PM1]{PM1}
\end{itemize}

\newpage

\chapter{IR: Infrastructure and services}

\section{IR1: Provide version control infrastructure for software and related research outputs to support FAIR practices through change tracking, provenance, and collaboration.}
\label{sec:IR1}

\subsection*{Description}
\begin{itemize}
  \item Motivation: Version control systems are indispensable tools for modern collaborative research. They enable multiple researchers to work on the same project simultaneously. This also helps in reducing duplications by enabling code reusability and improving researchers efficiency. This greatly facilitates knowledge transfer within and across research groups.
  \item Description: Version control systems for software and related research outputs—such as code, documentation, analysis workflows, and configuration management—enable tracking of changes, provenance maintenance, automation of testing processes, and effective collaboration among researchers.
  \item Description: The host laboratory should establish a clear timeline for support, gather feedback from experiments on requirements, and ensure that repositories can be ported to alternative systems if support is not extended and that any decisions made to modify repositories during such a transition are well documented.
  \item Description: In the case of national laboratories, national-level data infrastructures and services may be an option if the host laboratory does not provide such services.
\end{itemize}

\subsection*{Class}
\begin{itemize}
  \item Infrastructure and services
\end{itemize}

\subsection*{Actors}
\begin{itemize}
  \item host laboratory
\end{itemize}

\subsection*{Dependencies}
\begin{itemize}
  \item \hyperref[sec:CF2]{CF2}
  \item \hyperref[sec:CF1]{CF1}
\end{itemize}

\subsection*{Enables}
\begin{itemize}
  \item \hyperref[sec:CS2]{CS2}
  \item \hyperref[sec:CF6]{CF6}
  \item \hyperref[sec:SW9]{SW9}
  \item \hyperref[sec:AP6]{AP6}
  \item \hyperref[sec:SW3]{SW3}
  \item \hyperref[sec:SW2]{SW2}
  \item \hyperref[sec:SW1]{SW1}
  \item \hyperref[sec:CS3]{CS3}
  \item \hyperref[sec:AP7]{AP7}
\end{itemize}

\newpage

\section{IR2: Provide archiving infrastructure and support—including DOI assignment capabilities—for analysis-specific software, documentation, computational environments, and analysis workflows (as covered in the analysis preservation recommendations AP), to ensure proper citation, discoverability, and long-term accessibility.}
\label{sec:IR2}

\subsection*{Description}
\begin{itemize}
  \item Motivation: A major barrier to making analysis-specific software publicly available is the lack of a widely accepted repository infrastructure that meets the community’s needs and standards. A domain-specific infrastructure for software publication and archival would facilitate making analysis-specific code available.
  \item Description: The host laboratory should consider providing local infrastructure, such as an open-source digital repository with integrated FAIR and DOI minting capabilities. This strategy offers significant advantages in data governance, customization, and seamless integration with existing laboratory systems. Careful planning and appropriate resource allocation should be undertaken to address the associated efforts in infrastructure provision, maintenance, and DOI registration, maximizing the benefits for the laboratory and its community.
  \item Description: To maximize discoverability and interoperability, machine- and human-readable metadata should be defined to accompany deposited material.
  \item Description: For national laboratories—or in cases where a local solution is not feasible—national-level data infrastructures and services may serve as an alternative, particularly if the host laboratory does not offer such services.
\end{itemize}

\subsection*{Class}
\begin{itemize}
  \item Infrastructure and services
\end{itemize}

\subsection*{Actors}
\begin{itemize}
  \item host laboratory
\end{itemize}

\subsection*{Dependencies}
\begin{itemize}
  \item \hyperref[sec:CF2]{CF2}
  \item \hyperref[sec:CF1]{CF1}
\end{itemize}

\subsection*{Enables}
\begin{itemize}
  \item \hyperref[sec:AP11]{AP11}
  \item \hyperref[sec:CF6]{CF6}
  \item \hyperref[sec:LC4]{LC4}
  \item \hyperref[sec:PM7]{PM7}
  \item \hyperref[sec:AP14]{AP14}
  \item \hyperref[sec:CF3]{CF3}
  \item \hyperref[sec:AP7]{AP7}
\end{itemize}

\newpage

\section{IR3: Provide robust infrastructure and specialized personnel to develop and sustain open data repositories—including DOI assignment capabilities—to ensure long-term discoverability, accessibility and proper citation beyond the lifetime of individual experiments.}
\label{sec:IR3}

\subsection*{Description}
\begin{itemize}
  \item Description: Open data requires stable, persistent infrastructure that extends beyond the lifespan of any single experiment. Host laboratories should provide the institutional continuity needed for long-term data preservation and access, ensuring valuable research outputs remain available. This should be achieved through institutional or consortium data repositories. The access to open data in these repositories should be as easy as possible, preferably including streaming and standard download protocols.
  \item Description: In the case of national laboratories, national-level data infrastructures and services may be an option if the host laboratory does not provide such services.
\end{itemize}

\subsection*{Class}
\begin{itemize}
  \item Infrastructure and services
\end{itemize}

\subsection*{Actors}
\begin{itemize}
  \item host laboratory
\end{itemize}

\subsection*{Dependencies}
\begin{itemize}
  \item \hyperref[sec:CF2]{CF2}
  \item \hyperref[sec:CF1]{CF1}
  \item \hyperref[sec:PM13]{PM13}
\end{itemize}

\subsection*{Enables}
\begin{itemize}
  \item \hyperref[sec:CF6]{CF6}
  \item \hyperref[sec:PM5]{PM5}
  \item \hyperref[sec:CF3]{CF3}
  \item \hyperref[sec:DM6]{DM6}
\end{itemize}

\newpage

\section{IR4: Allocate dedicated infrastructure and personnel to host and maintain collaboration information resources, such as websites, for long-term accessibility beyond the lifetime of the experiment.}
\label{sec:IR4}

\subsection*{Description}
\begin{itemize}
  \item Motivation: Collaboration websites often contain critical technical knowledge, such as analysis procedures, calibration methods, and software configurations, that may not be documented elsewhere. This information is vital for understanding and reusing preserved data. Without proper infrastructure and ongoing maintenance, these resources risk disappearing shortly after a collaboration ends, undermining the usability and scientific value of preserved data.
\end{itemize}

\subsection*{Class}
\begin{itemize}
  \item Infrastructure and services
\end{itemize}

\subsection*{Actors}
\begin{itemize}
  \item host laboratory
\end{itemize}

\subsection*{Dependencies}
\begin{itemize}
  \item \hyperref[sec:CF1]{CF1}
  \item \hyperref[sec:PM14]{PM14}
\end{itemize}

\subsection*{Enables}
\begin{itemize}
  \item \hyperref[sec:DK4]{DK4}
  \item \hyperref[sec:PM14]{PM14}
\end{itemize}

\newpage

\section{IR5: Work with experiments’ management to identify long-term needs and provide suitable infrastructure for hosting conditions databases and essential related services that support future data reuse.}
\label{sec:IR5}

\subsection*{Description}
\begin{itemize}
  \item Motivation: Conditions databases store critical information such as calibrations, detector settings, and environmental parameters needed to accurately interpret preserved data. Maintaining fully operational systems over many years can be costly and challenging to sustain. By consulting with experiment teams to assess scientific needs, laboratories can prioritize maintaining essential components while gradually phasing out less critical services. This approach keeps preserved data usable for future research without overburdening infrastructure and staffing.
\end{itemize}

\subsection*{Class}
\begin{itemize}
  \item Infrastructure and services
\end{itemize}

\subsection*{Actors}
\begin{itemize}
  \item host laboratory
\end{itemize}

\subsection*{Dependencies}
\begin{itemize}
  \item \hyperref[sec:CF1]{CF1}
\end{itemize}

\subsection*{Enables}
\begin{itemize}
  \item \hyperref[sec:DM3]{DM3}
\end{itemize}

\newpage

\chapter{SK: Software skills development}

\section{SK1: Define the software skills curriculum required for data analysis work.}
\label{sec:SK1}

\subsection*{Description}
\begin{itemize}
  \item Motivation: Defining a particle-physics specific software curriculum, such as the curricula developed by the HSF training initiative and EVERSE, provides an opportunity for everyone to build or update skills relevant to their data analysis work. It enables equal chances for students to acquire software skills, regardless of their computing background or the specific expertise in their home institute. Having up-to-date software curricula available saves time for supervisors and WG leaders and facilitates organizing training events.
  \item Description: The curriculum should also cover computing practices focusing on FAIR principles and Open Science and archival guidelines.
  \item Description: A list of available training modules, as well as additional modules identified as desirable, should be updated regularly. Examples of additional desirable training modules may include topics such as machine learning or GPU programming - skills that are not yet essential for every physicist, but are increasingly in demand.
  \item Description: Not all aspects of this curriculum need to be specific to particle physics or HEP; if there are good external resources available for training, they should be leveraged rather than duplicated.
\end{itemize}

\subsection*{Class}
\begin{itemize}
  \item Software skills development
\end{itemize}

\subsection*{Actors}
\begin{itemize}
  \item experiment management
  \item WG leaders
\end{itemize}

\subsection*{Dependencies}
\begin{itemize}
  \item \hyperref[sec:SK4]{SK4}
\end{itemize}

\subsection*{Enables}
\begin{itemize}
  \item \hyperref[sec:SK2]{SK2}
  \item \hyperref[sec:SK6]{SK6}
  \item \hyperref[sec:SK3]{SK3}
  \item \hyperref[sec:CF5]{CF5}
  \item \hyperref[sec:SK7]{SK7}
\end{itemize}

\newpage

\section{SK2: As part of standard employee training and skill-building opportunities, allocate time in employees’ work plans for relevant software skills development.}
\label{sec:SK2}

\subsection*{Description}
\begin{itemize}
  \item Motivation: Adopting new skills requires time and it should not be expected that university software and computing courses are sufficient, nor that an adequate level of skill can be acquired just by “learning while doing”.
  \item Description: Home institutes should stay informed about available software skills curricula and relevant training initiatives. Given high personnel turnover and the trend toward careers in data-related industries, training should include standard tools that are widely used both inside and outside academia.
\end{itemize}

\subsection*{Class}
\begin{itemize}
  \item Software skills development
\end{itemize}

\subsection*{Actors}
\begin{itemize}
  \item home institute
\end{itemize}

\subsection*{Dependencies}
\begin{itemize}
  \item \hyperref[sec:SK1]{SK1}
\end{itemize}

\subsection*{Enables}
\begin{itemize}
  \item \hyperref[sec:PM9]{PM9}
  \item \hyperref[sec:SK6]{SK6}
  \item \hyperref[sec:SK3]{SK3}
  \item \hyperref[sec:AP2]{AP2}
  \item \hyperref[sec:CF5]{CF5}
  \item \hyperref[sec:SK5]{SK5}
\end{itemize}

\newpage

\section{SK3: Promote software skill development during employee onboarding and through early-career mentoring, schools, and workshops.}
\label{sec:SK3}

\subsection*{Description}
\begin{itemize}
  \item Motivation: Integrating skill development early ensures that new team members quickly acquire the competencies needed for effective data analysis and collaboration. Embedding training in onboarding and early-career programs helps reduce knowledge gaps, accelerate productivity, and foster a culture of continuous learning within the experiment.
\end{itemize}

\subsection*{Class}
\begin{itemize}
  \item Software skills development
\end{itemize}

\subsection*{Actors}
\begin{itemize}
  \item experiment management
  \item home institute
\end{itemize}

\subsection*{Dependencies}
\begin{itemize}
  \item \hyperref[sec:SK1]{SK1}
  \item \hyperref[sec:SK2]{SK2}
\end{itemize}

\subsection*{Enables}
\begin{itemize}
  \item \hyperref[sec:PM9]{PM9}
  \item \hyperref[sec:SW9]{SW9}
  \item \hyperref[sec:SW6]{SW6}
  \item \hyperref[sec:CF5]{CF5}
  \item \hyperref[sec:SW3]{SW3}
\end{itemize}

\newpage

\section{SK4: Actively support employees’ or group members’ contributions to community training initiatives in software skills, open science practices, and data management.}
\label{sec:SK4}

\subsection*{Description}
\begin{itemize}
  \item Motivation: Training in software skills, open science practices and data management relies on the expertise of practitioners who understand evolving tools, practices, and challenges. By actively supporting contributions to community-led training, institutions strengthen collective knowledge, enhance cross-domain practices, and ensure that training remains grounded in real-world experience.
\end{itemize}

\subsection*{Class}
\begin{itemize}
  \item Software skills development
\end{itemize}

\subsection*{Actors}
\begin{itemize}
  \item experiment management
  \item host laboratory
  \item home institute
  \item WG leaders
\end{itemize}

\subsection*{Enables}
\begin{itemize}
  \item \hyperref[sec:SK6]{SK6}
  \item \hyperref[sec:SK1]{SK1}
  \item \hyperref[sec:CF5]{CF5}
  \item \hyperref[sec:PM9]{PM9}
\end{itemize}

\newpage

\section{SK5: Include involvement in software skill training as a criterion in performance evaluations, with corresponding recognition or rewards.}
\label{sec:SK5}

\subsection*{Description}
\begin{itemize}
  \item Motivation: Many training initiatives have been developed within and for the community by experts in the field. To maintain the momentum of such initiatives, it is important to provide mechanisms for recognizing the work of trainers, mentors, and facilitators of training events.
  \item Motivation: Explicitly including involvement in software skills training in evaluation guidelines helps ensure these important activities are valued and widely adopted.
\end{itemize}

\subsection*{Class}
\begin{itemize}
  \item Software skills development
\end{itemize}

\subsection*{Actors}
\begin{itemize}
  \item experiment management
  \item home institute
\end{itemize}

\subsection*{Dependencies}
\begin{itemize}
  \item \hyperref[sec:SK2]{SK2}
\end{itemize}

\subsection*{Enables}
\begin{itemize}
  \item \hyperref[sec:CF5]{CF5}
  \item \hyperref[sec:PM9]{PM9}
\end{itemize}

\newpage

\section{SK6: Require project proposals to include software training plans focused on data and code management best practices that address preservation, documentation, and sharing.}
\label{sec:SK6}

\subsection*{Description}
\begin{itemize}
  \item Motivation: Software skills are essential for modern research, but many researchers lack formal training in software engineering, version control, data management, and workflow automation. At the same time, research software, including analysis code and tools, is valuable output that needs to be preserved and shared. Requiring training plans ensures projects build core competencies while following best practices for openness, reproducibility, and long-term access. This creates a research culture where software is recognized and supported as an essential component of scientific work.
\end{itemize}

\subsection*{Class}
\begin{itemize}
  \item Software skills development
\end{itemize}

\subsection*{Actors}
\begin{itemize}
  \item funding agency
\end{itemize}

\subsection*{Dependencies}
\begin{itemize}
  \item \hyperref[sec:SK2]{SK2}
  \item \hyperref[sec:PM7]{PM7}
  \item \hyperref[sec:SK1]{SK1}
  \item \hyperref[sec:AP14]{AP14}
  \item \hyperref[sec:AP15]{AP15}
  \item \hyperref[sec:SK4]{SK4}
\end{itemize}

\newpage

\section{SK7: Propose to and discuss with the relevant faculty or curriculum committee the possibility of awarding academic credit for research software skills training, regardless of whether these activities take place within or outside the formal curriculum.}
\label{sec:SK7}

\subsection*{Description}
\begin{itemize}
  \item Motivation: Awarding study credits for research software skills training provides direct motivation for students to engage in these activities and helps build a recognized training base—whether general, domain-specific, or experiment-specific—across universities and national borders.
\end{itemize}

\subsection*{Class}
\begin{itemize}
  \item Software skills development
\end{itemize}

\subsection*{Actors}
\begin{itemize}
  \item home institute
\end{itemize}

\subsection*{Dependencies}
\begin{itemize}
  \item \hyperref[sec:SK1]{SK1}
\end{itemize}

\newpage

\section{SK8: Promote the integration of dedicated software training courses and modules into undergraduate and graduate university curricula.}
\label{sec:SK8}

\subsection*{Description}
\begin{itemize}
  \item Motivation: Students joining experiments often lack foundational training in software and computing, which must then be addressed through supplementary initiatives at their home institutes. Explicitly providing structured training opportunities within university curricula will help close these gaps before students transition to research environments.
  \item Description: Research institutes that do not offer courses are encouraged to maintain a list of publicly available or self-study courses that students and employees can take advantage of (c.f. \hyperref[sec:SK2]{SK2}).
\end{itemize}

\subsection*{Class}
\begin{itemize}
  \item Software skills development
\end{itemize}

\subsection*{Actors}
\begin{itemize}
  \item home institute
\end{itemize}

\newpage

\chapter{LC: Licenses, copyright and citations}

\section{LC1: Mark intellectual property ownership with a copyright statement in accordance with host laboratory and home institute rules.}
\label{sec:LC1}

\subsection*{Description}
\begin{itemize}
  \item Description: A copyright notice is a brief statement that asserts the ownership of a work, and the time of its creation. It should indicate the copyright holder, and the year of creation of the element of your work that holds this copyright statement, such as a source file. Copyright ownership varies by institutional context and should be clearly indicated.
  \item Description: The copyright statement should be clearly specified, possibly using human and machine readable standards (for example SPDX, \url{https://spdx.dev/learn/handling-license-info/}), and should reflect the actual legal framework governing the work.
  \item Description: Experimental collaborations sometimes create specific copyright rules through memoranda of understanding. These should be understood and adopted by developers where appropriate.
\end{itemize}

\subsection*{Class}
\begin{itemize}
  \item Licenses, copyright and citations
\end{itemize}

\subsection*{Actors}
\begin{itemize}
  \item tool developers
  \item experiment management
  \item analysts
\end{itemize}

\subsection*{Enables}
\begin{itemize}
  \item \hyperref[sec:LC4]{LC4}
  \item \hyperref[sec:LC2]{LC2}
\end{itemize}

\newpage

\section{LC2: Apply a license approved by the Open Source Initiative (OSI) to your software.}
\label{sec:LC2}

\subsection*{Description}
\begin{itemize}
  \item Motivation: Using an Open Source Initiative-approved license provides clear legal terms for how software can be used, modified, and shared by the community. Merely making code available does not grant permission for reuse; only an explicit license ensures that others are legally allowed to access, use, and build upon the software. Description: A license should be applied to community-wide and collaboration-specific software tools and frameworks, but also to analysis-specific software.
  \item Description: The license should be consistent with the tooling environment, for example, license expectations in the scientific Python ecosystem. When combining and distributing software packages, make sure that licenses are compatible (inbound and outbound licenses in technical terms). Tools exist to assist the compatibility check, for example \url{https://interoperable-europe.ec.europa.eu/collection/eupl/solution/licensing-assistant}
  \item Description: License information should be clearly specified, possibly using human and machine readable standards (for example SPDX, \url{https://spdx.dev/learn/handling-license-info/}).
\end{itemize}

\subsection*{Class}
\begin{itemize}
  \item Licenses, copyright and citations
\end{itemize}

\subsection*{Actors}
\begin{itemize}
  \item tool developers
  \item analysts
\end{itemize}

\subsection*{Dependencies}
\begin{itemize}
  \item \hyperref[sec:LC1]{LC1}
\end{itemize}

\subsection*{Enables}
\begin{itemize}
  \item \hyperref[sec:LC4]{LC4}
  \item \hyperref[sec:AP14]{AP14}
  \item \hyperref[sec:CM1]{CM1}
\end{itemize}

\newpage

\section{LC3: Outline common guidelines for the citation of software and research data.}
\label{sec:LC3}

\subsection*{Description}
\begin{itemize}
  \item Motivation: Establishing common citation guidelines helps ensure consistent and proper acknowledgment of software and data across the collaboration. It facilitates the work of analysts, publication coordinators, and working group leaders by reducing ambiguity and providing clear expectations for what should be cited and how.
  \item Description: The guidelines should clarify which software elements should be cited in any relevant paper (e.g., specific software packages that enable a key analysis technique; software for which providers have requested citation). They should also clarify software groups that are general and need not be cited in every paper (e.g., Linux operating systems; ROOT for papers not focused on the details of ROOT file formats). The guidelines should be based on or refer to existing best practices such as \url{https://www.software.ac.uk/publication/how-cite-and-describe-software.}
  \item Description: The guidelines should specify that papers should contain a data availability statement and if applicable a proper reference to the dataset (with a DOI and adequate metadata).
\end{itemize}

\subsection*{Class}
\begin{itemize}
  \item Licenses, copyright and citations
\end{itemize}

\subsection*{Actors}
\begin{itemize}
  \item experiment management
\end{itemize}

\subsection*{Enables}
\begin{itemize}
  \item \hyperref[sec:LC5]{LC5}
  \item \hyperref[sec:LC6]{LC6}
\end{itemize}

\newpage

\section{LC4: Provide methods to make software products citable.}
\label{sec:LC4}

\subsection*{Description}
\begin{itemize}
  \item Motivation: Enabling citation of software itself—such as code repositories and releases—significantly increases the recognition and traceability of software contributions. Citation is only possible when clear guidance is provided; simply making code available is not sufficient. Dedicated citation metadata ensure users know exactly how to reference the software, facilitating proper attribution in research outputs.
  \item Description: To make software citable, include human- and machine-readable citation metadata in your repository using standardized formats such as CITATION.cff files. The Citation File Format (CFF) enables users and platforms to display and process citation information automatically.
\end{itemize}

\subsection*{Class}
\begin{itemize}
  \item Licenses, copyright and citations
\end{itemize}

\subsection*{Actors}
\begin{itemize}
  \item tool developers
\end{itemize}

\subsection*{Dependencies}
\begin{itemize}
  \item \hyperref[sec:LC2]{LC2}
  \item \hyperref[sec:IR2]{IR2}
  \item \hyperref[sec:LC1]{LC1}
\end{itemize}

\subsection*{Enables}
\begin{itemize}
  \item \hyperref[sec:LC5]{LC5}
\end{itemize}

\newpage

\section{LC5: Give priority to open-source software and tools with appropriate licences, and cite software appropriately in publications.}
\label{sec:LC5}

\subsection*{Description}
\begin{itemize}
  \item Motivation: Software based on non-open-source components cannot be freely shared, reused, or built upon, which limits reproducibility and long-term accessibility. Furthermore, relying on proprietary software complicates long-term preservation, especially if companies close or funding for collaborations stops.
  \item Description: Follow the experiment’s guideline for software citations as well as the citation requests of software providers, whether they specify a paper or a code repository. A proper citation is particularly important for open-source software projects, which rely on identifiable citations to demonstrate their usefulness in funding requests.
  \item Description: If no clear citation mechanisms exist for certain software tools, explore other means of giving credit to the software developers. Documenting the usage in code repositories in particular can be more explicit and inclusive than citations in eventual analysis papers.
\end{itemize}

\subsection*{Class}
\begin{itemize}
  \item Licenses, copyright and citations
\end{itemize}

\subsection*{Actors}
\begin{itemize}
  \item experiment management
  \item tool developers
  \item analysts
\end{itemize}

\subsection*{Dependencies}
\begin{itemize}
  \item \hyperref[sec:LC4]{LC4}
  \item \hyperref[sec:LC3]{LC3}
\end{itemize}

\newpage

\section{LC6: When using open data in publications, cite the data as specified in the corresponding open data record.}
\label{sec:LC6}

\subsection*{Description}
\begin{itemize}
  \item Motivation: Properly citing open data enables the impact and reuse of datasets to be monitored. Citation is only effective if followed in the recommended format; generic or incomplete references may prevent tracking and proper attribution of data use.
\end{itemize}

\subsection*{Class}
\begin{itemize}
  \item Licenses, copyright and citations
\end{itemize}

\subsection*{Actors}
\begin{itemize}
  \item tool developers
  \item analysts
\end{itemize}

\subsection*{Dependencies}
\begin{itemize}
  \item \hyperref[sec:LC3]{LC3}
\end{itemize}

\newpage

\chapter{CM: Community-wide software development}

\section{CM1: Ensure that the source code of community-wide software remains available in a trusted software archival repository.}
\label{sec:CM1}

\subsection*{Description}
\begin{itemize}
  \item Motivation: The long-term use and understanding of preserved data may depend on the tools available at the time of their initial use. Proper archival ensures that, as technology and computing environments evolve, the original code remains accessible for reference, re-execution, or adaptation.
\end{itemize}

\subsection*{Class}
\begin{itemize}
  \item Community-wide software development
\end{itemize}

\subsection*{Actors}
\begin{itemize}
  \item tool developers
\end{itemize}

\subsection*{Dependencies}
\begin{itemize}
  \item \hyperref[sec:LC2]{LC2}
\end{itemize}

\newpage

\section{CM2: Provide container images for community-wide software to enable consistent deployment and reproducible environments across heterogeneous computing platforms.}
\label{sec:CM2}

\subsection*{Description}
\begin{itemize}
  \item Motivation: Providing container images for community-wide software greatly simplifies their use in automated workflows and data analysis by ensuring consistent, reproducible environments across different computing platforms. This reduces installation and configuration barriers, saves time, and enables researchers to run analyses on open datasets or as part of larger automated pipelines. Typical tools included in such containers are ROOT, XRootD, and common Python packages.
  \item Description: The container images should follow the Open Container Initiative (OCI) compliance (\url{https://opencontainers.org/about/overview/}).
\end{itemize}

\subsection*{Class}
\begin{itemize}
  \item Community-wide software development
\end{itemize}

\subsection*{Actors}
\begin{itemize}
  \item tool developers
\end{itemize}

\subsection*{Dependencies}
\begin{itemize}
  \item \hyperref[sec:DK1]{DK1}
  \item \hyperref[sec:DK6]{DK6}
\end{itemize}

\newpage

\section{CM3: Maintain backward compatibility with earlier data formats by providing long-term read-only support and migration tools, and documenting changes affecting data access.}
\label{sec:CM3}

\subsection*{Description}
\begin{itemize}
  \item Motivation: Maintaining backward compatibility allows continued access to and reuse of preserved data.
\end{itemize}

\subsection*{Class}
\begin{itemize}
  \item Community-wide software development
\end{itemize}

\subsection*{Actors}
\begin{itemize}
  \item tool developers
\end{itemize}

\newpage

\section{CM4: Provide comprehensive documentation that corresponds to all available source code releases.}
\label{sec:CM4}

\subsection*{Description}
\begin{itemize}
  \item Motivation: To ensure long-term usability of preserved data, thorough documentation of the software tools needed in their analysis and guidance on their use are necessary.
\end{itemize}

\subsection*{Class}
\begin{itemize}
  \item Community-wide software development
\end{itemize}

\subsection*{Actors}
\begin{itemize}
  \item tool developers
\end{itemize}

\newpage

\chapter{CS: Collaboration-specific software development}

\section{CS1: Establish policies, procedures, and responsibilities for code review of widely-used analysis framework code.}
\label{sec:CS1}

\subsection*{Description}
\begin{itemize}
  \item Motivation: Code review ensures the quality and consistency of code contributions. General recommendations can be provided by the experiment management, while the working group leaders should be responsible for concrete recommendations based on the details of the developer community and software in question.
\end{itemize}

\subsection*{Class}
\begin{itemize}
  \item Collaboration-specific software development
\end{itemize}

\subsection*{Actors}
\begin{itemize}
  \item experiment management
  \item WG leaders
  \item tool developers
\end{itemize}

\subsection*{Dependencies}
\begin{itemize}
  \item \hyperref[sec:DK6]{DK6}
\end{itemize}

\newpage

\section{CS2: Provide clear “how-to-contribute” instruction templates for repositories, including guidance on repository collaboration tools such as branches, commit messages, merge/pull request roles, code review procedures, issue tracking, category tags, and milestones.}
\label{sec:CS2}

\subsection*{Description}
\begin{itemize}
  \item Motivation: How-to-contribute instructions encourage contributions from other collaborators whose knowledge may improve the analysis software. They also encourage people without prior expertise to contribute and build their software skills.
  \item Description: An example of how-to-contribute instructions \url{https://github.com/cernopendata/opendata.cern.ch/blob/master/CONTRIBUTING.rst}
\end{itemize}

\subsection*{Class}
\begin{itemize}
  \item Collaboration-specific software development
\end{itemize}

\subsection*{Actors}
\begin{itemize}
  \item WG leaders
  \item tool developers
\end{itemize}

\subsection*{Dependencies}
\begin{itemize}
  \item \hyperref[sec:IR1]{IR1}
  \item \hyperref[sec:DK6]{DK6}
\end{itemize}

\subsection*{Enables}
\begin{itemize}
  \item \hyperref[sec:AP5]{AP5}
\end{itemize}

\newpage

\section{CS3: Ensure that the source code for all versions of the central data-taking, simulation, and reconstruction software of the experiment is preserved, accessible and operable.}
\label{sec:CS3}

\subsection*{Description}
\begin{itemize}
  \item Motivation: Having the source code available allows for complete traceability of the data processing workflow, including the details of the reconstruction algorithms and physics object definitions. The code serves as the most accurate documentation of how data was handled, supporting future efforts to interpret and analyze the experiment’s data. Preserving the corresponding running environment—such as through software container images—further ensures that the code remains usable and executable in the future.
\end{itemize}

\subsection*{Class}
\begin{itemize}
  \item Collaboration-specific software development
\end{itemize}

\subsection*{Actors}
\begin{itemize}
  \item experiment management
  \item tool developers
\end{itemize}

\subsection*{Dependencies}
\begin{itemize}
  \item \hyperref[sec:IR1]{IR1}
\end{itemize}

\subsection*{Enables}
\begin{itemize}
  \item \hyperref[sec:DM5]{DM5}
  \item \hyperref[sec:CS6]{CS6}
  \item \hyperref[sec:DM9]{DM9}
  \item \hyperref[sec:DM7]{DM7}
\end{itemize}

\newpage

\section{CS4: Ensure that all collaboration-specific software not included in central dataset processing—such as code for deriving physics object corrections —is preserved, accessible, and properly documented with respect to the use of its outputs.}
\label{sec:CS4}

\subsection*{Description}
\begin{itemize}
  \item Motivation: After the standard dataset processing, additional derived quantities are often required for physics analyses. These are typically produced using specialized software developed by working groups or individual collaborators. Preserving this software and its documentation is essential to ensure that the entire data analysis workflow—including these important components—remains reproducible.
  \item Description: The preservation of this software should follow the principles established for overall analysis preservation, including the use of version control, appropriate dependencies, and long-term accessibility.
\end{itemize}

\subsection*{Class}
\begin{itemize}
  \item Collaboration-specific software development
\end{itemize}

\subsection*{Actors}
\begin{itemize}
  \item experiment management
  \item tool developers
\end{itemize}

\newpage

\section{CS5: Provide comprehensive documentation that corresponds to all available source code releases.}
\label{sec:CS5}

\subsection*{Description}
\begin{itemize}
  \item Motivation: Understanding how preserved data were processed requires accurate documentation of the processing software and workflows, including those used in earlier stages of an experiment. To ensure the correct and reproducible analysis of preserved data, thorough documentation of analysis software—and guidance on how to use it—is essential. Including documentation for past software versions ensures that legacy data can be interpreted and reused even as tools evolve over time.
\end{itemize}

\subsection*{Class}
\begin{itemize}
  \item Collaboration-specific software development
\end{itemize}

\subsection*{Actors}
\begin{itemize}
  \item experiment management
  \item tool developers
\end{itemize}

\newpage

\section{CS6: Promote the use of documentation building tools and provide training on their use.}
\label{sec:CS6}

\subsection*{Description}
\begin{itemize}
  \item Motivation: Such tools simplify documentation generation and make this process more collaborative, therefore potentially more useful. Collaboration/experiment wide use of such tools promote a culture of documentation development and maintenance on every level (e.g. core software, analysis software, etc.). Clear and cross-referenced software documentation benefits current and future developers and users.
\end{itemize}

\subsection*{Class}
\begin{itemize}
  \item Collaboration-specific software development
\end{itemize}

\subsection*{Actors}
\begin{itemize}
  \item experiment management
  \item tool developers
\end{itemize}

\subsection*{Dependencies}
\begin{itemize}
  \item \hyperref[sec:CS3]{CS3}
\end{itemize}

\newpage

\chapter{SW: Software and workflow management - analysis-specific SW}

\section{SW1: Mandate the use of findable and accessible version control infrastructure for analysis-specific software.}
\label{sec:SW1}

\subsection*{Description}
\begin{itemize}
  \item Motivation: All software used within the experiment should be findable and accessible to all members of the experiment. Experience shows that this is hard to achieve without defining a common code-hosting solution for all analyses and making its use mandatory.
  \item Motivation: Deciding on and enforcing the use of a common version control infrastructure for analysis code within the experiment makes analysis code findable and accessible to all different analysis groups, also in the long term. This facilitates the use of common tools and sharing knowledge.
  \item Description: Clearly indicate the common version control infrastructure to be used for all analysis code, and establish a preferred software license for analysis code repositories to be included by default when repositories are created.
\end{itemize}

\subsection*{Class}
\begin{itemize}
  \item Software and workflow management - analysis-specific SW
\end{itemize}

\subsection*{Actors}
\begin{itemize}
  \item experiment management
\end{itemize}

\subsection*{Dependencies}
\begin{itemize}
  \item \hyperref[sec:IR1]{IR1}
\end{itemize}

\subsection*{Enables}
\begin{itemize}
  \item \hyperref[sec:AP5]{AP5}
  \item \hyperref[sec:AP6]{AP6}
  \item \hyperref[sec:AP3]{AP3}
  \item \hyperref[sec:SW4]{SW4}
  \item \hyperref[sec:SW2]{SW2}
\end{itemize}

\newpage

\section{SW2: From the early stages of an analysis, store your code in findable and accessible repositories within the version control infrastructure designated by the experiment.}
\label{sec:SW2}

\subsection*{Description}
\begin{itemize}
  \item Motivation: Common analysis code repositories within an experiment make analysis code findable and accessible to all members of the experiment. This facilitates the use of common tools, sharing knowledge, and transparent analysis review.
\end{itemize}

\subsection*{Class}
\begin{itemize}
  \item Software and workflow management - analysis-specific SW
\end{itemize}

\subsection*{Actors}
\begin{itemize}
  \item analysts
\end{itemize}

\subsection*{Dependencies}
\begin{itemize}
  \item \hyperref[sec:SW1]{SW1}
  \item \hyperref[sec:IR1]{IR1}
\end{itemize}

\subsection*{Enables}
\begin{itemize}
  \item \hyperref[sec:SW9]{SW9}
  \item \hyperref[sec:SW11]{SW11}
  \item \hyperref[sec:SW8]{SW8}
\end{itemize}

\newpage

\section{SW3: Use code versioning tools (e.g. Git), and commit changes frequently to the common repository.}
\label{sec:SW3}

\subsection*{Description}
\begin{itemize}
  \item Motivation: Code versioning ensures a clear history of changes, and enables easy rollback to previous versions. Frequent commits help catch issues early, make your progress visible, and reduce conflicts by integrating small, manageable updates regularly.
\end{itemize}

\subsection*{Class}
\begin{itemize}
  \item Software and workflow management - analysis-specific SW
\end{itemize}

\subsection*{Actors}
\begin{itemize}
  \item analysts
\end{itemize}

\subsection*{Dependencies}
\begin{itemize}
  \item \hyperref[sec:SK3]{SK3}
  \item \hyperref[sec:IR1]{IR1}
\end{itemize}

\subsection*{Enables}
\begin{itemize}
  \item \hyperref[sec:SW4]{SW4}
  \item \hyperref[sec:SW11]{SW11}
  \item \hyperref[sec:AP7]{AP7}
\end{itemize}

\newpage

\section{SW4: Use, where available and appropriate, repository collaboration tools such as issue tracking and merge/pull requests, category tags, and milestones.}
\label{sec:SW4}

\subsection*{Description}
\begin{itemize}
  \item Motivation: Issue tracking provides documentation of problems and solutions found during the software development lifecycle. Merge/pull requests and their reviews document the changes to the source code. Use of category tags and milestones helps to categorize issues and organize work around specific goals. All of these tools aid in collaborative software development and provide documentation for present and future developers.
\end{itemize}

\subsection*{Class}
\begin{itemize}
  \item Software and workflow management - analysis-specific SW
\end{itemize}

\subsection*{Actors}
\begin{itemize}
  \item analysts
\end{itemize}

\subsection*{Dependencies}
\begin{itemize}
  \item \hyperref[sec:SW3]{SW3}
  \item \hyperref[sec:SW1]{SW1}
\end{itemize}

\subsection*{Enables}
\begin{itemize}
  \item \hyperref[sec:AP14]{AP14}
\end{itemize}

\newpage

\section{SW5: If contributions to an analysis code repository from other developers are welcomed, provide clear “how-to-contribute” instructions.}
\label{sec:SW5}

\subsection*{Description}
\begin{itemize}
  \item Motivation: How-to-contribute instructions encourage contributions from other collaborators whose knowledge may improve the analysis software. They also encourage people without prior expertise to contribute and build their software skills.
  \item Description: The instructions should indicate use of issue tracking, branches, commit messages, roles for merging commits, reviewing code. An example of how-to-contribute instructions \url{https://github.com/cernopendata/opendata.cern.ch/blob/master/CONTRIBUTING.rst}
\end{itemize}

\subsection*{Class}
\begin{itemize}
  \item Software and workflow management - analysis-specific SW
\end{itemize}

\subsection*{Actors}
\begin{itemize}
  \item WG leaders
  \item analysts
\end{itemize}

\subsection*{Dependencies}
\begin{itemize}
  \item \hyperref[sec:DK1]{DK1}
\end{itemize}

\subsection*{Enables}
\begin{itemize}
  \item \hyperref[sec:AP5]{AP5}
  \item \hyperref[sec:AP14]{AP14}
\end{itemize}

\newpage

\section{SW6: Avoid hardcoding input parameters that are likely to change within the lifecycle of your analysis or for similar analyses in the future; make the code configurable instead.}
\label{sec:SW6}

\subsection*{Description}
\begin{itemize}
  \item Motivation: Configurable input parameters facilitate reuse of the code or parts of it within the lifecycle of your analysis or for similar analyses in the future. This flexibility allows adjustment of parameters without changing the code and helps distinguish input parameters from the code itself, improving readability and maintainability.
  \item Description: For configurable scripts, provide whatever analysis-specific run commands are required to replicate the analysis procedure (e.g. a shell script, a workflow implementation or similar).
\end{itemize}

\subsection*{Class}
\begin{itemize}
  \item Software and workflow management - analysis-specific SW
\end{itemize}

\subsection*{Actors}
\begin{itemize}
  \item analysts
\end{itemize}

\subsection*{Dependencies}
\begin{itemize}
  \item \hyperref[sec:SK3]{SK3}
  \item \hyperref[sec:DK1]{DK1}
\end{itemize}

\newpage

\section{SW7: Structure the analysis code into repositories or subdirectories that correspond to different steps in the analysis workflow.}
\label{sec:SW7}

\subsection*{Description}
\begin{itemize}
  \item Motivation: Structuring the analysis code into repositories or subdirectories that correspond to different steps in the analysis workflow improves readability and makes the code well adapted for automated workflows.
  \item Description: Where needed, provide clear links to other code repositories on which certain analysis steps depend.
\end{itemize}

\subsection*{Class}
\begin{itemize}
  \item Software and workflow management - analysis-specific SW
\end{itemize}

\subsection*{Actors}
\begin{itemize}
  \item analysts
\end{itemize}

\subsection*{Enables}
\begin{itemize}
  \item \hyperref[sec:SW9]{SW9}
  \item \hyperref[sec:AP12]{AP12}
  \item \hyperref[sec:SW10]{SW10}
\end{itemize}

\newpage

\section{SW8: Use well-defined environments, such as software containers or virtual environments, and make them explicit in the code repository.}
\label{sec:SW8}

\subsection*{Description}
\begin{itemize}
  \item Motivation: Using well-defined environments makes it much easier to share code with others and to run analyses reliably in automated workflows, since all necessary dependencies are explicitly specified and reproducible.
  \item Description: Where software is reasonably lightweight, an effort should be made to make these environments entirely stand-alone, without external dependencies. This is important for future reuse scenarios where central infrastructure is no longer available (e.g. calibration files or metadata distributed via central files systems such as CVMFS at CERN).
\end{itemize}

\subsection*{Class}
\begin{itemize}
  \item Software and workflow management - analysis-specific SW
\end{itemize}

\subsection*{Actors}
\begin{itemize}
  \item analysts
\end{itemize}

\subsection*{Dependencies}
\begin{itemize}
  \item \hyperref[sec:AP3]{AP3}
  \item \hyperref[sec:SW2]{SW2}
  \item \hyperref[sec:DK1]{DK1}
  \item \hyperref[sec:AP13]{AP13}
\end{itemize}

\subsection*{Enables}
\begin{itemize}
  \item \hyperref[sec:SW9]{SW9}
  \item \hyperref[sec:AP14]{AP14}
\end{itemize}

\newpage

\section{SW9: Implement meaningful automated tests at multiple levels to ensure code quality and facilitate reproducible analyses.}
\label{sec:SW9}

\subsection*{Description}
\begin{itemize}
  \item Motivation: Automated tests check the functionality at different levels, e.g. code compiling, producing expected output at different analysis workflow steps. User experience in analysis groups has shown that adopting proper testing, i.e. CI/CD pipelines, has greatly improved the working efficiency in the group since changes to the analysis code are tested immediately.
  \item Motivation: Setting up a testing sequence also sets up an analysis workflow that uses a well-defined computing environment. This helps detect when the analysis code does not work beyond the analysts’ local computing environment, allowing issues to be identified and corrected.
\end{itemize}

\subsection*{Class}
\begin{itemize}
  \item Software and workflow management - analysis-specific SW
\end{itemize}

\subsection*{Actors}
\begin{itemize}
  \item analysts
  \item tool developers
\end{itemize}

\subsection*{Dependencies}
\begin{itemize}
  \item \hyperref[sec:SW7]{SW7}
  \item \hyperref[sec:SK3]{SK3}
  \item \hyperref[sec:AP3]{AP3}
  \item \hyperref[sec:SW2]{SW2}
  \item \hyperref[sec:IR1]{IR1}
  \item \hyperref[sec:SW8]{SW8}
\end{itemize}

\newpage

\section{SW10: Implement and use well-defined software workflows, and provide detailed comments and/or dedicated documentation.}
\label{sec:SW10}

\subsection*{Description}
\begin{itemize}
  \item Motivation: Having the code available helps understand and document single steps in the workflow, but understanding the analysis workflow requires further documentation or using proper workflow languages (e.g. snakemake…). Meaningful automated tests implemented as a CI/CD pipeline can also clarify the sequence of the workflow.
\end{itemize}

\subsection*{Class}
\begin{itemize}
  \item Software and workflow management - analysis-specific SW
\end{itemize}

\subsection*{Actors}
\begin{itemize}
  \item analysts
\end{itemize}

\subsection*{Dependencies}
\begin{itemize}
  \item \hyperref[sec:DK1]{DK1}
  \item \hyperref[sec:SW7]{SW7}
\end{itemize}

\subsection*{Enables}
\begin{itemize}
  \item \hyperref[sec:AP10]{AP10}
  \item \hyperref[sec:AP12]{AP12}
  \item \hyperref[sec:AP14]{AP14}
\end{itemize}

\newpage

\section{SW11: Create checkpoints of analysis code repositories, for example through Git tags, to clearly correspond to the stages of the analysis.}
\label{sec:SW11}

\subsection*{Description}
\begin{itemize}
  \item Motivation: These checkpoints provide clear reference points at a given stage of the work. They are helpful if you need to retrace your steps, fix a bug, or justify a decision. Unlike branches, which are often used for ongoing, parallel development, tags act as permanent markers for important states of the code. They simplify code rollbacks, if needed, and are useful for version comparisons to evaluate new features.
  \item Description: Working group leaders: incorporate requests for code repository tags into review procedures to promote their usage.
\end{itemize}

\subsection*{Class}
\begin{itemize}
  \item Software and workflow management - analysis-specific SW
\end{itemize}

\subsection*{Actors}
\begin{itemize}
  \item WG leaders
  \item analysts
\end{itemize}

\subsection*{Dependencies}
\begin{itemize}
  \item \hyperref[sec:SW3]{SW3}
  \item \hyperref[sec:SW2]{SW2}
\end{itemize}

\subsection*{Enables}
\begin{itemize}
  \item \hyperref[sec:AP14]{AP14}
  \item \hyperref[sec:AP7]{AP7}
\end{itemize}

\newpage

\section{SW12: Prioritize contributions to established software infrastructure over the development of new tools in funding guidelines and evaluations.}
\label{sec:SW12}

\subsection*{Description}
\begin{itemize}
  \item Motivation: Developing new software tools from scratch often duplicates existing capabilities and results in unsustainable standalone projects that become obsolete after funding ends. By prioritizing contributions to existing infrastructure, funding agencies promote efficient resource use, enhance software sustainability, and strengthen the overall research software ecosystem through collaboration.
\end{itemize}

\subsection*{Class}
\begin{itemize}
  \item Software and workflow management - analysis-specific SW
\end{itemize}

\subsection*{Actors}
\begin{itemize}
  \item funding agency
\end{itemize}

\subsection*{Enables}
\begin{itemize}
  \item \hyperref[sec:AP5]{AP5}
\end{itemize}

\newpage

\chapter{AP: Analysis preservation tools and practices}

\section{AP1: Provide tools and storage for internal analysis information, and enforce their use.}
\label{sec:AP1}

\subsection*{Description}
\begin{itemize}
  \item Motivation: Central tools make this information easily findable and searchable within the experimental collaboration, which is especially important given the large size and long duration of these collaborations. Programmatic access to this information enables queries such as determining overall dataset usage or identifying the exact input dataset for a given publication. It also enables assessing e.g. the usage of simulated datasets, improves their findability within the experiment, and can be used to enrich the information about them once made public.
  \item Description: This information can include dataset details (such as input dataset lists, formats, and versions used), any relevant metadata (e.g. keywords, data taking periods, statistical tools, methodology), analysis metadata (membership, relevant dates, contact information), and repository information (code and paper repositories in use). All information should be stored in a format that is both human- and machine-readable, with programmatic access available.
\end{itemize}

\subsection*{Class}
\begin{itemize}
  \item Analysis preservation tools and practices
\end{itemize}

\subsection*{Actors}
\begin{itemize}
  \item experiment management
\end{itemize}

\subsection*{Enables}
\begin{itemize}
  \item \hyperref[sec:AP4]{AP4}
  \item \hyperref[sec:AP13]{AP13}
\end{itemize}

\newpage

\section{AP2: Encourage the adoption of best practices (SW2-11) starting from the early stages of an analysis.}
\label{sec:AP2}

\subsection*{Description}
\begin{itemize}
  \item Motivation: Adopting best practices early allows analysts to become familiar and comfortable with these standards as an integral part of their work, rather than viewing them as external requirements imposed later in the process.
  \item Motivation: If the expected outcome of best practices is only enforced at the review stage, analysts may struggle to provide them, and opportunities for learning and skills improvement are lost.
  \item Description: Support analysts by providing relevant examples, training, and hands-on mentoring.
\end{itemize}

\subsection*{Class}
\begin{itemize}
  \item Analysis preservation tools and practices
\end{itemize}

\subsection*{Actors}
\begin{itemize}
  \item experiment management
  \item WG leaders
\end{itemize}

\subsection*{Dependencies}
\begin{itemize}
  \item \hyperref[sec:SK2]{SK2}
  \item \hyperref[sec:PM8]{PM8}
\end{itemize}

\subsection*{Enables}
\begin{itemize}
  \item \hyperref[sec:AP11]{AP11}
  \item \hyperref[sec:AP14]{AP14}
  \item \hyperref[sec:PM9]{PM9}
\end{itemize}

\newpage

\section{AP3: Provide common testing infrastructure and testing templates.}
\label{sec:AP3}

\subsection*{Description}
\begin{itemize}
  \item Motivation: Common testing templates, such as code compilation and container building in automated CI/CD pipelines, speed up adoption of good coding practices.
  \item Description: Experiments can provide container images with the most common software available, as well as mechanisms to access experiment- or host-laboratory-specific infrastructure. For example, containers can provide common disk areas and software (or data) distributions via CVMFS (a network file system widely used in high energy physics to distribute software and data across sites). The choice of distribution method should match the experiment's computing infrastructure and institutional capabilities.
\end{itemize}

\subsection*{Class}
\begin{itemize}
  \item Analysis preservation tools and practices
\end{itemize}

\subsection*{Actors}
\begin{itemize}
  \item tool developers
  \item WG leaders
\end{itemize}

\subsection*{Dependencies}
\begin{itemize}
  \item \hyperref[sec:PM3]{PM3}
  \item \hyperref[sec:SW1]{SW1}
  \item \hyperref[sec:PM8]{PM8}
\end{itemize}

\subsection*{Enables}
\begin{itemize}
  \item \hyperref[sec:SW9]{SW9}
  \item \hyperref[sec:AP6]{AP6}
  \item \hyperref[sec:SW8]{SW8}
\end{itemize}

\newpage

\section{AP4: Record and store analysis configuration information in human- and machine-readable form.}
\label{sec:AP4}

\subsection*{Description}
\begin{itemize}
  \item Motivation: Saving details such as dataset names, trigger paths, and other parameter values, and storing them in commonly accessible locations, makes it easy to check and reuse the exact inputs from previous analyses.
  \item Description: This information (e.g. input dataset names, trigger paths etc) should be collected in an experiment-internal database, or another searchable service.
\end{itemize}

\subsection*{Class}
\begin{itemize}
  \item Analysis preservation tools and practices
\end{itemize}

\subsection*{Actors}
\begin{itemize}
  \item analysts
\end{itemize}

\subsection*{Dependencies}
\begin{itemize}
  \item \hyperref[sec:AP1]{AP1}
\end{itemize}

\subsection*{Enables}
\begin{itemize}
  \item \hyperref[sec:AP9]{AP9}
  \item \hyperref[sec:PM15]{PM15}
\end{itemize}

\newpage

\section{AP5: Encourage the use of, and contributions to, common analysis tools.}
\label{sec:AP5}

\subsection*{Description}
\begin{itemize}
  \item Motivation: Encouraging the use of and contribution to common analysis tools improves work efficiency for analysis groups, increases the maintainability of code, and supports collaboration between analysis teams.
  \item Description: Common analysis tools include processing tools acting directly on centrally processed datasets, orchestration tools for managing physics analysis workflows, tools designed to facilitate the long-term reproducibility of analyses, and commonly used statistical analysis tools. Promoting the adoption of such tools helps reduce the risks associated with the proliferation of analysis frameworks, such as duplicated functionality, increased maintenance burdens, and the persistence of unmaintained legacy code.
  \item Description: Also consider the reasons why framework proliferation occurs, such as difficulty contributing to common projects (due to a lack of findability, clear instructions, or incompatible development timescales) or a preference for working in smaller, dynamic groups. To alleviate these issues, improve the discoverability and documentation of common tools, provide clear contribution guidelines, and encourage broader participation in development teams so that they can support flexible development practices.
\end{itemize}

\subsection*{Class}
\begin{itemize}
  \item Analysis preservation tools and practices
\end{itemize}

\subsection*{Actors}
\begin{itemize}
  \item experiment management
  \item WG leaders
\end{itemize}

\subsection*{Dependencies}
\begin{itemize}
  \item \hyperref[sec:CS2]{CS2}
  \item \hyperref[sec:PM8]{PM8}
  \item \hyperref[sec:SW5]{SW5}
  \item \hyperref[sec:SW1]{SW1}
  \item \hyperref[sec:SW12]{SW12}
\end{itemize}

\subsection*{Enables}
\begin{itemize}
  \item \hyperref[sec:PM9]{PM9}
\end{itemize}

\newpage

\section{AP6: Integrate automated analysis workflows into analysis software or common software tools, and encourage their implementation across working groups and experiments.}
\label{sec:AP6}

\subsection*{Description}
\begin{itemize}
  \item Motivation: New tools and methods for workflow automation are continually emerging, but they may not be widely known or adopted across all groups. Regular integration of these methods into new tools and analysis software helps to build a culture of automation and thinking about preservation and reusability. Regular group or experiment-level activities promoting automation help share experience, knowledge, and tools between teams, making it easier to identify and implement improvements.
  \item Description: Analysts and tool developers: proactively develop tools and methods to ease and automate analysis workflows and, if applicable, integrate them in a common set of tools for wider use.
  \item Description: WG leaders and experiment management: encourage implementation of automation by planning regular initiatives.
\end{itemize}

\subsection*{Class}
\begin{itemize}
  \item Analysis preservation tools and practices
\end{itemize}

\subsection*{Actors}
\begin{itemize}
  \item experiment management
  \item WG leaders
  \item tool developers
  \item analysts
\end{itemize}

\subsection*{Dependencies}
\begin{itemize}
  \item \hyperref[sec:AP3]{AP3}
  \item \hyperref[sec:SW1]{SW1}
  \item \hyperref[sec:PM8]{PM8}
  \item \hyperref[sec:IR1]{IR1}
\end{itemize}

\subsection*{Enables}
\begin{itemize}
  \item \hyperref[sec:AP11]{AP11}
  \item \hyperref[sec:PM9]{PM9}
  \item \hyperref[sec:AP9]{AP9}
  \item \hyperref[sec:AP12]{AP12}
  \item \hyperref[sec:AP10]{AP10}
\end{itemize}

\newpage

\section{AP7: Require that code in analysis code repositories be checkpointed using a version-control tag or archived to correspond to the publication.}
\label{sec:AP7}

\subsection*{Description}
\begin{itemize}
  \item Motivation: A checkpoint or an archival repository provides a permanent reference to the exact version of the code used for the published results and makes it possible to openly share software as a research product. This serves as an unambiguous and complete documentation of the analysis process.
\end{itemize}

\subsection*{Class}
\begin{itemize}
  \item Analysis preservation tools and practices
\end{itemize}

\subsection*{Actors}
\begin{itemize}
  \item experiment management
  \item WG leaders
\end{itemize}

\subsection*{Dependencies}
\begin{itemize}
  \item \hyperref[sec:SW3]{SW3}
  \item \hyperref[sec:SW11]{SW11}
  \item \hyperref[sec:IR1]{IR1}
  \item \hyperref[sec:IR2]{IR2}
\end{itemize}

\subsection*{Enables}
\begin{itemize}
  \item \hyperref[sec:AP8]{AP8}
  \item \hyperref[sec:AP12]{AP12}
  \item \hyperref[sec:AP14]{AP14}
\end{itemize}

\newpage

\section{AP8: Require that all code and configuration necessary to produce the final plots in an analysis be findable, available, and reusable as a condition for approval.}
\label{sec:AP8}

\subsection*{Description}
\begin{itemize}
  \item Motivation: As team members leave a project, it becomes increasingly difficult to recover or reconstruct the code and input configurations necessary to reproduce analysis plots after the analysis has been submitted for approval. Making findability, availability, and reusability of this material a requirement at the time of approval should be considered a minimal standard for preserving analysis results, and represents an important first step toward safeguarding analysis-specific software and workflows. This measure is also essential because, if plots are stored solely as images, they cannot be contributed to repositories such as HEPData, which require results in a machine-readable format.
  \item Description: Establish this as an explicit requirement during the approval process, and define clear criteria and procedures to verify that the relevant code and configurations are findable, available, and reusable. At this point, "reusability" specifically means that collaboration members can readily rerun, or update the final stage of the analysis to regenerate the final results if necessary.
\end{itemize}

\subsection*{Class}
\begin{itemize}
  \item Analysis preservation tools and practices
\end{itemize}

\subsection*{Actors}
\begin{itemize}
  \item experiment management
  \item WG leaders
\end{itemize}

\subsection*{Dependencies}
\begin{itemize}
  \item \hyperref[sec:AP7]{AP7}
\end{itemize}

\subsection*{Enables}
\begin{itemize}
  \item \hyperref[sec:AP9]{AP9}
  \item \hyperref[sec:AP12]{AP12}
  \item \hyperref[sec:PM9]{PM9}
\end{itemize}

\newpage

\section{AP9: Make the analysis code and configuration information needed for final plots and results findable, available, and reusable.}
\label{sec:AP9}

\subsection*{Description}
\begin{itemize}
  \item Motivation: Ensuring that the code and inputs used to produce approved results and final plots are findable, available, and reusable allows others to reproduce these results and update them if needed. Organizing and preserving these materials throughout the analysis process helps protect results from being lost as team members move on or projects evolve.
  \item Description: At the final stage of the analysis, all code, inputs and configuration information required to generate the final results and plots should be organized, documented, and stored such that they are easily accessible within the collaboration. At this point, "reusability" specifically means that collaboration members can readily rerun or update the final stage of the analysis to regenerate the final results if necessary.
  \item Description: Wherever possible, implement this requirement using automated workflows that package code, configuration files, and related documentation together as part of the analysis finalization process.
\end{itemize}

\subsection*{Class}
\begin{itemize}
  \item Analysis preservation tools and practices
\end{itemize}

\subsection*{Actors}
\begin{itemize}
  \item analysts
\end{itemize}

\subsection*{Dependencies}
\begin{itemize}
  \item \hyperref[sec:AP6]{AP6}
  \item \hyperref[sec:AP4]{AP4}
  \item \hyperref[sec:AP12]{AP12}
  \item \hyperref[sec:AP8]{AP8}
\end{itemize}

\newpage

\section{AP10: Provide selected reusable analysis workflows that are compatible with the released or forthcoming open data.}
\label{sec:AP10}

\subsection*{Description}
\begin{itemize}
  \item Motivation: Planning for reusability during active analysis makes it possible to provide research-quality usage examples after the data have been released to the public.
  \item Description: A good practice is to select representative analyses covering a range of topics, and to allocate personnel to adapt these workflows to the preserved data and public documentation. Full replicability should not be expected, as the released data may have undergone different processing than the original analysis data.
\end{itemize}

\subsection*{Class}
\begin{itemize}
  \item Analysis preservation tools and practices
\end{itemize}

\subsection*{Actors}
\begin{itemize}
  \item experiment management
  \item WG leaders
  \item analysts
\end{itemize}

\subsection*{Dependencies}
\begin{itemize}
  \item \hyperref[sec:DK1]{DK1}
  \item \hyperref[sec:AP6]{AP6}
  \item \hyperref[sec:SW10]{SW10}
\end{itemize}

\newpage

\section{AP11: Set a goal to establish reproducible analyses as a standard, and provide clear, actionable guidelines covering code, software environments, workflows, metadata, and documentation.}
\label{sec:AP11}

\subsection*{Description}
\begin{itemize}
  \item Motivation: Clear and actionable guidelines for reproducible analyses help analysts make code and workflows—now recognized as fundamental research outputs—accessible, understandable, and usable by others. This is an emerging best practice crucial for open science, providing an unambiguous and complete documentation of the analysis process and supporting long-term preservation and future usability of preserved data.
  \item Motivation: By adopting these guidelines, the analysis process within the collaboration—with access to the input data—becomes FAIR (Findable, Accessible, Interoperable, and Reusable), ensuring that analyses can be efficiently located, understood, and reused by collaborators, and laying the groundwork for extending these benefits when data and materials are released more broadly.
  \item Motivation: Analysis code, workflow and environment descriptions, and documentation can then be made public at the time of publication of results and be assigned a DOI, improving their findability, accessibility, and academic recognition. To maximize discoverability and interoperability, these guidelines should also define machine- and human-readable metadata that should accompany analysis code and workflows.
\end{itemize}

\subsection*{Class}
\begin{itemize}
  \item Analysis preservation tools and practices
\end{itemize}

\subsection*{Actors}
\begin{itemize}
  \item experiment management
  \item WG leaders
\end{itemize}

\subsection*{Dependencies}
\begin{itemize}
  \item \hyperref[sec:AP2]{AP2}
  \item \hyperref[sec:AP6]{AP6}
  \item \hyperref[sec:IR2]{IR2}
  \item \hyperref[sec:PM7]{PM7}
\end{itemize}

\subsection*{Enables}
\begin{itemize}
  \item \hyperref[sec:AP12]{AP12}
  \item \hyperref[sec:PM9]{PM9}
\end{itemize}

\newpage

\section{AP12: Document or package the software environments used to produce the final analysis results.}
\label{sec:AP12}

\subsection*{Description}
\begin{itemize}
  \item Motivation: Providing the full software environment for each analysis workflow step (including library versions, for example providing a container image or a requirements file) saves considerable effort during code reuse and ensures your analysis results can be reliably reproduced in different environments. This saves time, reduces errors, and makes it possible to extend on these results in the future.
\end{itemize}

\subsection*{Class}
\begin{itemize}
  \item Analysis preservation tools and practices
\end{itemize}

\subsection*{Actors}
\begin{itemize}
  \item analysts
\end{itemize}

\subsection*{Dependencies}
\begin{itemize}
  \item \hyperref[sec:AP11]{AP11}
  \item \hyperref[sec:AP8]{AP8}
  \item \hyperref[sec:SW7]{SW7}
  \item \hyperref[sec:AP6]{AP6}
  \item \hyperref[sec:SW10]{SW10}
  \item \hyperref[sec:DK1]{DK1}
  \item \hyperref[sec:AP7]{AP7}
\end{itemize}

\subsection*{Enables}
\begin{itemize}
  \item \hyperref[sec:AP9]{AP9}
  \item \hyperref[sec:AP14]{AP14}
\end{itemize}

\newpage

\section{AP13: Provide analysis metadata and supplementary data products, such as corrections, in centrally managed, version-controlled locations so that they can be linked unambiguously to an analysis.}
\label{sec:AP13}

\subsection*{Description}
\begin{itemize}
  \item Motivation: Central, version-controlled storage ensures that the exact data, corrections, and any other supplementary data products used in an analysis are always identifiable. These records need to be maintained for the entire lifetime of the experiment’s data analysis and preserved for future use.
\end{itemize}

\subsection*{Class}
\begin{itemize}
  \item Analysis preservation tools and practices
\end{itemize}

\subsection*{Actors}
\begin{itemize}
  \item experiment management
  \item WG leaders
  \item analysts
\end{itemize}

\subsection*{Dependencies}
\begin{itemize}
  \item \hyperref[sec:AP1]{AP1}
\end{itemize}

\subsection*{Enables}
\begin{itemize}
  \item \hyperref[sec:DM3]{DM3}
  \item \hyperref[sec:SW8]{SW8}
\end{itemize}

\newpage

\section{AP14: Publish analysis-specific code and workflow descriptions to archival repositories at the time of the paper publication.  for DOI generation.}
\label{sec:AP14}

\subsection*{Description}
\begin{itemize}
  \item Motivation: For open science, code should be considered an integral part of the research outcome. Publishing analysis-specific code and workflow descriptions in archival repositories with a DOI is an emerging best practice that enables full transparency of research methods, ensures the reusability of code, and allows the application of these workflows to future public data releases.
  \item Description: Following good software practices during analysis makes this process manageable. Code versions should be checkpointed to capture the precise state used to produce published results. Workflow descriptions should also be published to clearly and unambiguously capture each step of the analysis process, enabling others to understand and follow the full sequence of operations. Documentation should be clear and sufficiently detailed to enable someone unfamiliar with the work to understand the code and workflows, but remain focused on essential information to avoid unnecessary effort.
\end{itemize}

\subsection*{Class}
\begin{itemize}
  \item Analysis preservation tools and practices
\end{itemize}

\subsection*{Actors}
\begin{itemize}
  \item WG leaders
  \item analysts
\end{itemize}

\subsection*{Dependencies}
\begin{itemize}
  \item \hyperref[sec:IR2]{IR2}
  \item \hyperref[sec:SW5]{SW5}
  \item \hyperref[sec:SW10]{SW10}
  \item \hyperref[sec:PM7]{PM7}
  \item \hyperref[sec:AP2]{AP2}
  \item \hyperref[sec:AP12]{AP12}
  \item \hyperref[sec:LC2]{LC2}
  \item \hyperref[sec:AP15]{AP15}
  \item \hyperref[sec:DK1]{DK1}
  \item \hyperref[sec:SW4]{SW4}
  \item \hyperref[sec:SW11]{SW11}
  \item \hyperref[sec:SW8]{SW8}
  \item \hyperref[sec:AP7]{AP7}
\end{itemize}

\subsection*{Enables}
\begin{itemize}
  \item \hyperref[sec:SK6]{SK6}
\end{itemize}

\newpage

\section{AP15: Require preservation and public sharing of analysis software and research methods as a condition of funding.}
\label{sec:AP15}

\subsection*{Description}
\begin{itemize}
  \item Motivation: Funding agency guidelines can play a crucial role in advancing open science by setting clear expectations and standards for the openness of research methods through the preservation of analysis software and knowledge related to funded projects. By requiring preservation and public sharing of analysis software and methods, funding agencies promote the adoption of these practices in everyday research work and ensure that research outcomes are reusable and accessible to others.
  \item Description: The mechanism for sharing, where appropriate, should match the standards for the collaboration or experiment. That is, software and data from a specific grant should be integrated into and made public as a part of the collaboration’s or experiment’s general efforts, and not either independently or using different infrastructure.
\end{itemize}

\subsection*{Class}
\begin{itemize}
  \item Analysis preservation tools and practices
\end{itemize}

\subsection*{Actors}
\begin{itemize}
  \item funding agency
\end{itemize}

\subsection*{Enables}
\begin{itemize}
  \item \hyperref[sec:SK6]{SK6}
  \item \hyperref[sec:AP14]{AP14}
\end{itemize}

\newpage

\chapter{DM: Data management tools and practices}

\section{DM1: For each data-taking period, complete and document the final (“legacy”) processing version of the collected data and the corresponding simulations.}
\label{sec:DM1}

\subsection*{Description}
\begin{itemize}
  \item Motivation: Defining a legacy processing for each data-taking period specifies exactly which datasets should be used when combining data from different periods in new analyses. If new data are collected, a new legacy processing round can be established to ensure all data are processed consistently for future combined analyses.
  \item Description: The definition of the legacy processing should include a complete list of datasets and clear instructions on how the datasets and their metadata can be queried from the experiment’s databases. It should also specify the software versions and any additional data required to analyze these datasets.
\end{itemize}

\subsection*{Class}
\begin{itemize}
  \item Data management tools and practices
\end{itemize}

\subsection*{Actors}
\begin{itemize}
  \item data management
\end{itemize}

\subsection*{Enables}
\begin{itemize}
  \item \hyperref[sec:DM9]{DM9}
\end{itemize}

\newpage

\section{DM2: Ensure that the main data management tools used to extract supplementary data—such as online event selection information and data processing workflows—needed for proper analysis and understanding of preserved data remain accessible.}
\label{sec:DM2}

\subsection*{Description}
\begin{itemize}
  \item Motivation: Data management tools evolve over time, but it is essential to ensure continued access to metadata such as online event selections and processing workflows.
  \item Description: If older versions of these tools cannot be kept accessible, the information they contain should be exported to a standard, readable open format and properly archived. This guarantees long-term access to essential metadata for interpreting and reusing preserved data.
\end{itemize}

\subsection*{Class}
\begin{itemize}
  \item Data management tools and practices
\end{itemize}

\subsection*{Actors}
\begin{itemize}
  \item data management
\end{itemize}

\subsection*{Enables}
\begin{itemize}
  \item \hyperref[sec:DM9]{DM9}
  \item \hyperref[sec:DM7]{DM7}
\end{itemize}

\newpage

\section{DM3: Ensure the long-term availability of supplementary data and other key information necessary for the reusability of preserved data.}
\label{sec:DM3}

\subsection*{Description}
\begin{itemize}
  \item Motivation: The long-term research value of preserved data depends on availability of various supplementary information, such as luminosity details, data quality information, corrections, scale factors, and condition data. Without these, researchers in the future cannot fully interpret or reliably reuse the data. Experience from previous experiments has shown that even file catalogs - which are essential for locating and managing datasets - can be lost over time, so special care should be taken to preserve them as well.
\end{itemize}

\subsection*{Class}
\begin{itemize}
  \item Data management tools and practices
\end{itemize}

\subsection*{Actors}
\begin{itemize}
  \item data management
\end{itemize}

\subsection*{Dependencies}
\begin{itemize}
  \item \hyperref[sec:IR5]{IR5}
  \item \hyperref[sec:AP13]{AP13}
\end{itemize}

\subsection*{Enables}
\begin{itemize}
  \item \hyperref[sec:DM9]{DM9}
  \item \hyperref[sec:DM5]{DM5}
  \item \hyperref[sec:DM7]{DM7}
\end{itemize}

\newpage

\section{DM4: Ensure long-term storage of data with sufficient resilience to protect against site-specific risks, with a preference for maintaining more than one copy in geographically distinct locations when this is practical and motivated by the importance of the datasets.}
\label{sec:DM4}

\subsection*{Description}
\begin{itemize}
  \item Motivation: Single sites are subject to rare catastrophic events that could put at risk data at the site (floods, fires, etc).
  \item Description: Data should be distributed over at least two sites. If practical, at least two full copies of each data file should be distributed geographically. As long as the simulation workflows and software are maintained, the simulated datasets do not need to be treated with the same extreme care applied to the detector data. Inter-laboratory collaboration could provide a model by which small experiments’ data are stored in another location, even when collaborators are geographically centralized.
\end{itemize}

\subsection*{Class}
\begin{itemize}
  \item Data management tools and practices
\end{itemize}

\subsection*{Actors}
\begin{itemize}
  \item experiment management
  \item host laboratory
\end{itemize}

\subsection*{Dependencies}
\begin{itemize}
  \item \hyperref[sec:PM13]{PM13}
\end{itemize}

\newpage

\section{DM5: Record data processing information, including trigger details for collision data, generator parameters for simulated data, data processing steps (such as software versions, input configurations, and other runtime inputs) and other relevant metadata for all types of data.}
\label{sec:DM5}

\subsection*{Description}
\begin{itemize}
  \item Motivation: Processing information documents every step in the data processing chain and is essential for preserving the details of event selection and reconstruction algorithms.
\end{itemize}

\subsection*{Class}
\begin{itemize}
  \item Data management tools and practices
\end{itemize}

\subsection*{Actors}
\begin{itemize}
  \item data management
\end{itemize}

\subsection*{Dependencies}
\begin{itemize}
  \item \hyperref[sec:DM3]{DM3}
  \item \hyperref[sec:CS3]{CS3}
\end{itemize}

\subsection*{Enables}
\begin{itemize}
  \item \hyperref[sec:DM6]{DM6}
  \item \hyperref[sec:DM7]{DM7}
\end{itemize}

\newpage

\section{DM6: Publish event-level experimental data and simulations to repositories aligned with an experiment's open science policy and institutional requirements.}
\label{sec:DM6}

\subsection*{Description}
\begin{itemize}
  \item Motivation: Publishing event-level experimental data and simulations in suitable open repositories is the key element of open science in particle physics. It maximizes the return on investment in basic research by enabling the widest possible reuse and analysis of the data.
  \item Motivation: Unrestricted, authentication-free read access to open data eliminates barriers, ensuring that researchers and other users can efficiently discover, access, and utilize these resources without unnecessary hurdles.
\end{itemize}

\subsection*{Class}
\begin{itemize}
  \item Data management tools and practices
\end{itemize}

\subsection*{Actors}
\begin{itemize}
  \item experiment management
  \item open data group
\end{itemize}

\subsection*{Dependencies}
\begin{itemize}
  \item \hyperref[sec:PM5]{PM5}
  \item \hyperref[sec:PM6]{PM6}
  \item \hyperref[sec:DM5]{DM5}
  \item \hyperref[sec:IR3]{IR3}
  \item \hyperref[sec:DM10]{DM10}
  \item \hyperref[sec:DM7]{DM7}
\end{itemize}

\subsection*{Enables}
\begin{itemize}
  \item \hyperref[sec:DM8]{DM8}
\end{itemize}

\newpage

\section{DM7: Ensure proper metadata accompanies preserved datasets, and provide the software necessary for their use.}
\label{sec:DM7}

\subsection*{Description}
\begin{itemize}
  \item Motivation: Proper metadata makes datasets understandable, discoverable, and reusable. Three key areas should be covered: content metadata (such as dataset size and content type), processing metadata (data collection or generation mechanisms, data processing steps), and contextual metadata (instructions on how to use datasets in research contexts and how to utilize related supplementary data).
  \item Description: Metadata should be machine-readable and include basic data information, processing workflows, etc.
  \item Description: Software required to utilize preserved data should be accessible and well documented, including usage guidelines and analysis methods.
  \item Description: Design metadata elements according to experiment needs whilst taking into account their possible future use in contexts going beyond the current experiment scope.
\end{itemize}

\subsection*{Class}
\begin{itemize}
  \item Data management tools and practices
\end{itemize}

\subsection*{Actors}
\begin{itemize}
  \item data management
  \item open data group
\end{itemize}

\subsection*{Dependencies}
\begin{itemize}
  \item \hyperref[sec:DM2]{DM2}
  \item \hyperref[sec:DM5]{DM5}
  \item \hyperref[sec:DM3]{DM3}
  \item \hyperref[sec:CS3]{CS3}
\end{itemize}

\subsection*{Enables}
\begin{itemize}
  \item \hyperref[sec:DM9]{DM9}
  \item \hyperref[sec:DM6]{DM6}
\end{itemize}

\newpage

\section{DM8: Organise regular campaigns in which scientists outside the collaboration test the reproducibility of selected analyses.}
\label{sec:DM8}

\subsection*{Description}
\begin{itemize}
  \item Motivation: Providing tools and instructions for open data usage that are easy to understand and implement by external users is a challenge. Feedback from targeted user tests is invaluable to maintaining truly usable open data.
  \item Description: The testing should access data and simulations in the open data repositories and leverage the tools provided as part of the open science policy implementation.
\end{itemize}

\subsection*{Class}
\begin{itemize}
  \item Data management tools and practices
\end{itemize}

\subsection*{Actors}
\begin{itemize}
  \item experiment management
  \item open data group
\end{itemize}

\subsection*{Dependencies}
\begin{itemize}
  \item \hyperref[sec:DM6]{DM6}
  \item \hyperref[sec:DK7]{DK7}
\end{itemize}

\newpage

\section{DM9: Ensure that everything required to reproduce legacy data and to regenerate simulated data is available and functional in case of accidental data loss.}
\label{sec:DM9}

\subsection*{Description}
\begin{itemize}
  \item Motivation: If the data preservation strategy relies on keeping raw data with the assumption that all derived data can be regenerated from it, it is essential to verify and regularly test that the full reprocessing chain—from software and configurations to processing environments—is available and functional. If this is not the case, the legacy reprocessings themselves must be preserved with due care.
  \item Description: This includes preserving all necessary software, workflows, processing environments (such as container images or virtual machines), and any additional data or configuration files needed during processing. Regular testing ensures that legacy datasets can be fully restored if they are accidentally lost.
\end{itemize}

\subsection*{Class}
\begin{itemize}
  \item Data management tools and practices
\end{itemize}

\subsection*{Actors}
\begin{itemize}
  \item data management
\end{itemize}

\subsection*{Dependencies}
\begin{itemize}
  \item \hyperref[sec:DM1]{DM1}
  \item \hyperref[sec:DM2]{DM2}
  \item \hyperref[sec:DM3]{DM3}
  \item \hyperref[sec:DM7]{DM7}
  \item \hyperref[sec:CS3]{CS3}
\end{itemize}

\newpage

\section{DM10: Ensure datasets released publicly use standardized formats and are assigned persistent identifiers.}
\label{sec:DM10}

\subsection*{Description}
\begin{itemize}
  \item Motivation: Standardized formats (e.g. ROOT, \hyperref[sec:HDF5]{HDF5}) ensure both data portability and software compatibility. Persistent identifiers (e.g., DOIs) ensure reliable long-term access, proper attribution, and compliance with FAIR principles. Usage of standardized formats for open data will facilitate AI/ML utilization by making generation of training datasets more straightforward.
\end{itemize}

\subsection*{Class}
\begin{itemize}
  \item Data management tools and practices
\end{itemize}

\subsection*{Actors}
\begin{itemize}
  \item open data group
\end{itemize}

\subsection*{Enables}
\begin{itemize}
  \item \hyperref[sec:DM6]{DM6}
\end{itemize}

\newpage

\section{DM11: Whenever possible, structure data format content to be accessible and analyzable using only community-standard tools (e.g., the ROOT framework), without the need for experiment-specific software.}
\label{sec:DM11}

\subsection*{Description}
\begin{itemize}
  \item Motivation: Providing simpler data formats reduces long-term software dependencies, making it more likely that future researchers can use and analyze preserved data even if experiment-specific tools become obsolete. By relying on community-standard tools, data preservation and interoperability are improved, facilitating integration with new analysis workflows and compatibility with evolving technologies.
\end{itemize}

\subsection*{Class}
\begin{itemize}
  \item Data management tools and practices
\end{itemize}

\subsection*{Actors}
\begin{itemize}
  \item experiment management
  \item data management
\end{itemize}

\newpage

\chapter{DK: Documentation and knowledge preservation}

\section{DK1: Provide sufficient documentation in each software repository for other researchers to understand and use the code.}
\label{sec:DK1}

\subsection*{Description}
\begin{itemize}
  \item Motivation: Documentation helps newcomers and collaborators understand and use the code. It also supports the author during development and maintenance. Clear documentation is a fundamental part of software best practices.
\end{itemize}

\subsection*{Class}
\begin{itemize}
  \item Documentation and knowledge preservation
\end{itemize}

\subsection*{Actors}
\begin{itemize}
  \item analysts
\end{itemize}

\subsection*{Dependencies}
\begin{itemize}
  \item \hyperref[sec:DK2]{DK2}
\end{itemize}

\subsection*{Enables}
\begin{itemize}
  \item \hyperref[sec:SW10]{SW10}
  \item \hyperref[sec:SW5]{SW5}
  \item \hyperref[sec:SW6]{SW6}
  \item \hyperref[sec:AP12]{AP12}
  \item \hyperref[sec:AP14]{AP14}
  \item \hyperref[sec:CM2]{CM2}
  \item \hyperref[sec:AP10]{AP10}
  \item \hyperref[sec:SW8]{SW8}
\end{itemize}

\newpage

\section{DK2: Incorporate thorough documentation of work as a criterion in employee performance evaluations, with appropriate recognition or rewards.}
\label{sec:DK2}

\subsection*{Description}
\begin{itemize}
  \item Motivation: Proper documentation is often seen as extra work beyond an employee’s primary research tasks, but it is a critical part of good scientific practice. Recognizing and rewarding documentation in performance evaluations encourages researchers to treat it as an integral part of their work.
\end{itemize}

\subsection*{Class}
\begin{itemize}
  \item Documentation and knowledge preservation
\end{itemize}

\subsection*{Actors}
\begin{itemize}
  \item experiment management
  \item home institute
\end{itemize}

\subsection*{Enables}
\begin{itemize}
  \item \hyperref[sec:DK1]{DK1}
  \item \hyperref[sec:PM9]{PM9}
\end{itemize}

\newpage

\section{DK3: Avoid using proprietary or closed-source formats for documentation, instructions, and tutorials.}
\label{sec:DK3}

\subsection*{Description}
\begin{itemize}
  \item Motivation: Ensuring long-term accessibility and usability of documentation is essential for preserving knowledge and supporting future users. Binary formatted files may not be readable in 30 years. For long-term availability, simple and robust open-source formats should be preferred, so that content remains accessible regardless of the underlying implementation.
  \item Description: Even when using open-source formats, it is important to consider the supporting infrastructure—for example, while markdown is open, the system used to render it or certain modules might be closed-source.
  \item Description: When updating legacy documentation, efforts should be made to convert materials - such as collaboration policy documents that are often in proprietary software formats - into open-source formats.
\end{itemize}

\subsection*{Class}
\begin{itemize}
  \item Documentation and knowledge preservation
\end{itemize}

\subsection*{Actors}
\begin{itemize}
  \item analysts
  \item experiment management
  \item WG leaders
\end{itemize}

\newpage

\section{DK4: Establish procedures to guarantee that documentation stays accessible throughout the experiment's lifecycle and can be openly shared when appropriate, with clear protocols for preservation and access management beyond the experiment’s duration.}
\label{sec:DK4}

\subsection*{Description}
\begin{itemize}
  \item Motivation: Documentation often becomes inaccessible over time due to platform changes, broken links, obsolete formats, or authentication issues, endangering the preservation of essential technical knowledge. This approach ensures the continuity of institutional knowledge, supports future reuse, and promotes open access where possible.
  \item Description: Experiments and host laboratories should adopt documentation preservation strategies that ensure findability, include regular accessibility checks, and plan for format migration. Even if documentation cannot be publicly available during the experiment’s active phase, it should be preserved with controlled access for host laboratory personnel after collaboration ends. When authentication systems are decommissioned at project completion, clear protocols should outline how documentation access transitions to long-term systems.
\end{itemize}

\subsection*{Class}
\begin{itemize}
  \item Documentation and knowledge preservation
\end{itemize}

\subsection*{Actors}
\begin{itemize}
  \item experiment management
  \item host laboratory
\end{itemize}

\subsection*{Dependencies}
\begin{itemize}
  \item \hyperref[sec:IR4]{IR4}
  \item \hyperref[sec:PM14]{PM14}
  \item \hyperref[sec:PM1]{PM1}
\end{itemize}

\newpage

\section{DK5: Establish guidelines for making public relevant sections of internal notes and documentation.}
\label{sec:DK5}

\subsection*{Description}
\begin{itemize}
  \item Motivation: Internal notes often contain valuable technical information that may be important for future data reuse or legacy data usability. The goal is not meant to make documentation harder to write, but to remind authors that useful content might be shared publicly if it supports long-term access and understanding.
  \item Description: Clearly communicate to the collaboration that notes or portions thereof containing useful technical details may become public, so that authors can follow style and content guidelines suitable for publicly shared documents.
\end{itemize}

\subsection*{Class}
\begin{itemize}
  \item Documentation and knowledge preservation
\end{itemize}

\subsection*{Actors}
\begin{itemize}
  \item experiment management
  \item WG leaders
\end{itemize}

\newpage

\section{DK6: Encourage developers and maintainers of relevant software tool repositories to provide clear documentation on code design principles and contributing practices to facilitate easy and efficient onboarding of new team members.}
\label{sec:DK6}

\subsection*{Description}
\begin{itemize}
  \item Motivation: Clear documentation - separate from usage instructions - on code design principles and contributing practices ensures that new developers and maintainers can quickly understand the software tools and contribute effectively from the start. It also helps attract new contributors to the project. This information can be included as in-repository documentation where appropriate.
\end{itemize}

\subsection*{Class}
\begin{itemize}
  \item Documentation and knowledge preservation
\end{itemize}

\subsection*{Actors}
\begin{itemize}
  \item WG leaders
\end{itemize}

\subsection*{Enables}
\begin{itemize}
  \item \hyperref[sec:CM2]{CM2}
  \item \hyperref[sec:CS1]{CS1}
  \item \hyperref[sec:CS2]{CS2}
\end{itemize}

\newpage

\section{DK7: Accompany openly released data with example analysis workflows to increase their usability.}
\label{sec:DK7}

\subsection*{Description}
\begin{itemize}
  \item Motivation: Open datasets are most usable when accompanied by introductory example analysis workflows designed for the general scientific public. Simplified examples focus on demonstrating concepts in a clear way for non-experts in particle physics. Such examples complement more in-depth workflows designed to reproduce the results of a publication.
\end{itemize}

\subsection*{Class}
\begin{itemize}
  \item Documentation and knowledge preservation
\end{itemize}

\subsection*{Actors}
\begin{itemize}
  \item open data group
\end{itemize}

\subsection*{Enables}
\begin{itemize}
  \item \hyperref[sec:DM8]{DM8}
\end{itemize}

\newpage

\section{DK8: Through application guidelines and funding decisions, ensure that funded projects contributing to the experiments or utilizing their data provide sufficient documentation of their research outputs to guarantee their reusability.}
\label{sec:DK8}

\subsection*{Description}
\begin{itemize}
  \item Motivation: Funding agency guidelines can play a crucial role in advancing open science by setting clear expectations and standards on research outcome reusability. Comprehensive documentation, including proper metadata, is essential for the long-term reuse and understanding of research products. Good documentation and metadata support the application of the FAIR principles, enabling future researchers to interpret, validate, and extend previous work, and maximizing the value and impact of funded research.
\end{itemize}

\subsection*{Class}
\begin{itemize}
  \item Documentation and knowledge preservation
\end{itemize}

\subsection*{Actors}
\begin{itemize}
  \item funding agency
\end{itemize}

\newpage

\chapter{LS: Long-term sustainability}

\section{LS1: Ensure a proper long-term archiving ecosystem compliant with the OAIS (Open Archival Information System) reference model is in place, including professional archivists.}
\label{sec:LS1}

\subsection*{Description}
\begin{itemize}
  \item Motivation: Long-term in archiving means giving the best possible change for the archived material to be read in more than 100 years. Following the latest standards and good practices is a continuous effort.
\end{itemize}

\subsection*{Class}
\begin{itemize}
  \item Long-term sustainability
\end{itemize}

\subsection*{Actors}
\begin{itemize}
  \item host laboratory
\end{itemize}

\newpage

\section{LS2: Assign dedicated human, software, and hardware resources to monitor, maintain, and respond to changes in computing infrastructure that may affect access to and reusability of legacy data.}
\label{sec:LS2}

\subsection*{Description}
\begin{itemize}
  \item Motivation: Evolving computing infrastructure, such as changes in file systems, storage protocols, or container environments, can disrupt access to preserved data and break essential tools and workflows. Without proactive monitoring and maintenance, legacy data becomes increasingly vulnerable to obsolescence. Consistent infrastructure management ensures the long-term usability of data and safeguards the scientific value of past research investments.
  \item Description: Host laboratories must dedicate resources to test access to preserved data regularly, identify emerging compatibility risks, and implement timely updates or migrations. This includes retaining expertise in legacy systems, tracking relevant technology roadmaps, and preparing contingency plans for critical preservation workflows.
\end{itemize}

\subsection*{Class}
\begin{itemize}
  \item Long-term sustainability
\end{itemize}

\subsection*{Actors}
\begin{itemize}
  \item host laboratory
\end{itemize}

\subsection*{Dependencies}
\begin{itemize}
  \item \hyperref[sec:CF2]{CF2}
  \item \hyperref[sec:CF1]{CF1}
\end{itemize}

\newpage

\section{LS3: Maintain a registry of all external software dependencies required for preserved data and their reusability, and archive their source code and documentation in a persistent repository whenever feasible and permitted by licensing.}
\label{sec:LS3}

\subsection*{Description}
\begin{itemize}
  \item Description: If a key component disappears, the software stack may have to be adapted, or maintenance of the component needs to be insourced. (Example: CERNLIB)
  \item Description: In some cases, software license agreements may prohibit the preservation or distribution of external software in this way. Such cases should be identified by the experiment management and host laboratory, and explicit decisions should be made about the appropriate path forward that retains maximal scientific flexibility with minimal legal risk.
\end{itemize}

\subsection*{Class}
\begin{itemize}
  \item Long-term sustainability
\end{itemize}

\subsection*{Actors}
\begin{itemize}
  \item host laboratory
  \item experiment management
\end{itemize}

\newpage

\section{LS4: Ensure that changes to storage infrastructure do not put archived data at additional risk.}
\label{sec:LS4}

\subsection*{Description}
\begin{itemize}
  \item Motivation: Increasing media size can raise risk by concentrating more data on a single storage element. Sufficient effort must be allocated for any required data migration, with the host laboratory responsible for these tasks after the experiment ends.
  \item Description: Note that the technology evolution of the storage media market, e.g. the availability of tapes and their cost can increase pressure on funding resources.
\end{itemize}

\subsection*{Class}
\begin{itemize}
  \item Long-term sustainability
\end{itemize}

\subsection*{Actors}
\begin{itemize}
  \item host laboratory
  \item experiment management
\end{itemize}

\subsection*{Enables}
\begin{itemize}
  \item \hyperref[sec:PM13]{PM13}
\end{itemize}

\newpage

\section{LS5: Establish a technological watch process to monitor computing technology evolution and develop strategies that aim at long-term data and software accessibility.}
\label{sec:LS5}

\subsection*{Description}
\begin{itemize}
  \item Motivation: Rapid evolution in computing technologies, including hardware architectures, compilers, and programming languages, poses persistent risks to long-term data accessibility and software functionality. A structured technology watch process enables experiments and host institutions to anticipate disruptions across all technology layers and implement timely migration strategies, ensuring valuable scientific assets remain usable and citable over time.
  \item Using software container images to preserve the original computing environment can mitigate many of the impacts of technology evolution, such as compiler changes, programming language modifications, and system architecture updates. However, containers are not a permanent solution.  For example, not all containers will run smoothly on new architectures - some may simply not be compatible. To reduce this risk, it is advisable to adopt multi architecture builds for critical software early on and make gradual updates or adaptations as needed, so that essential applications remain usable as hardware evolves. Regular checks and updates are important to keep data and software accessible as technology changes.
\end{itemize}

\subsection*{Class}
\begin{itemize}
  \item Long-term sustainability
\end{itemize}

\subsection*{Actors}
\begin{itemize}
  \item experiment management
  \item host laboratory
\end{itemize}

\newpage

\chapter{CF: Cost, funding and return of investment}

\section{CF1: Establish dedicated, sustainable funding for data preservation and open science infrastructure, personnel, storage, and operations, supported by explicit budget line items and multi-year commitments.}
\label{sec:CF1}

\subsection*{Description}
\begin{itemize}
  \item Motivation: Explicit budgeting ensures that data preservation and open science are adequately resourced, covering infrastructure, operations, and dedicated personnel, formally integrated into the host laboratory’s organigram, rather than relying on voluntary or ad hoc contributions from staff with competing responsibilities. Sustained, multi-year funding is essential, as preservation efforts extend beyond the lifespan of individual experiments and require long-term investment in storage, computing, and technical expertise. Without explicit budgeting, these activities remain vulnerable to staff turnover and competing priorities, ultimately jeopardizing the long-term accessibility of preserved research outputs.
\end{itemize}

\subsection*{Class}
\begin{itemize}
  \item Cost, funding and return of investment
\end{itemize}

\subsection*{Actors}
\begin{itemize}
  \item host laboratory
  \item funding agency
\end{itemize}

\subsection*{Dependencies}
\begin{itemize}
  \item \hyperref[sec:CF2]{CF2}
  \item \hyperref[sec:PM2]{PM2}
\end{itemize}

\subsection*{Enables}
\begin{itemize}
  \item \hyperref[sec:CF6]{CF6}
  \item \hyperref[sec:IR2]{IR2}
  \item \hyperref[sec:PM9]{PM9}
  \item \hyperref[sec:IR5]{IR5}
  \item \hyperref[sec:IR3]{IR3}
  \item \hyperref[sec:IR4]{IR4}
  \item \hyperref[sec:IR1]{IR1}
  \item \hyperref[sec:LS2]{LS2}
\end{itemize}

\newpage

\section{CF2: Allocate funding for sustained data preservation and open science through dual mechanisms that provide experiment-level support during active phases and laboratory-level support for long-term maintenance and access beyond the lifetime of individual experiments.}
\label{sec:CF2}

\subsection*{Description}
\begin{itemize}
  \item Motivation: Data preservation and open science require sustained investment across both time and organizational levels. Experiment-level funding enables collaborations to prepare data for preservation, develop workflows, and establish infrastructure during active data-taking. Laboratory-level funding ensures long-term maintenance, access, and technology updates after project-specific funding ends. This dual approach reflects the reality that preservation costs extend well beyond the lifespan of individual experiments and cannot be fully supported by project budgets. Institutional funding commitments are essential to maintain infrastructure, support specialized personnel, and meet evolving technological needs. Without coordinated support at both levels, valuable research investments risk becoming inaccessible, undermining the long-term return on public research investments.
\end{itemize}

\subsection*{Class}
\begin{itemize}
  \item Cost, funding and return of investment
\end{itemize}

\subsection*{Actors}
\begin{itemize}
  \item funding agency
\end{itemize}

\subsection*{Dependencies}
\begin{itemize}
  \item \hyperref[sec:PM2]{PM2}
  \item \hyperref[sec:PM17]{PM17}
\end{itemize}

\subsection*{Enables}
\begin{itemize}
  \item \hyperref[sec:IR2]{IR2}
  \item \hyperref[sec:PM13]{PM13}
  \item \hyperref[sec:IR3]{IR3}
  \item \hyperref[sec:CF1]{CF1}
  \item \hyperref[sec:IR1]{IR1}
  \item \hyperref[sec:LS2]{LS2}
\end{itemize}

\newpage

\section{CF3: Establish comprehensive mechanisms to track and quantify the reuse of preserved data and knowledge through DOI citations, usage metrics, and regular community surveys, to demonstrate scientific and societal impact and guide future preservation strategies.}
\label{sec:CF3}

\subsection*{Description}
\begin{itemize}
  \item Motivation: Systematic tracking of data and knowledge reuse provides essential evidence of the scientific and societal value of preservation efforts, allowing for data-driven decisions on resource allocation and long-term strategies. A variety of indicators should be monitored—including DOI citations, dataset downloads, repository activity, documentation views, container pulls, and training material usage—to fully capture the impact across research, education, and outreach. Community surveys complement automated metrics by capturing less visible forms of reuse, such as classroom applications or unpublished ongoing research. This comprehensive approach promotes transparency and accountability, highlights successful practices, and identifies areas needing additional support, demonstrating return on investment to stakeholders and guiding more effective preservation planning.
\end{itemize}

\subsection*{Class}
\begin{itemize}
  \item Cost, funding and return of investment
\end{itemize}

\subsection*{Actors}
\begin{itemize}
  \item host laboratory
  \item experiment management
\end{itemize}

\subsection*{Dependencies}
\begin{itemize}
  \item \hyperref[sec:IR3]{IR3}
  \item \hyperref[sec:IR2]{IR2}
\end{itemize}

\newpage

\section{CF4: Keep track of (quantify) the continued analysis activity on preserved data within the experiment, particularly in the period towards the end of the lifetime of the experiment when no further reprocessing is planned.}
\label{sec:CF4}

\subsection*{Description}
\begin{itemize}
  \item Motivation: Quantifying ongoing analysis activity on preserved data demonstrates the continued scientific value and impact of data preservation efforts, especially as the experiment winds down. Tracking this usage provides clear evidence that investments in data preservation enable further research, maximize the return on resources spent, and support new discoveries even after active data collection and reprocessing have ended.
\end{itemize}

\subsection*{Class}
\begin{itemize}
  \item Cost, funding and return of investment
\end{itemize}

\subsection*{Actors}
\begin{itemize}
  \item experiment management
\end{itemize}

\newpage

\section{CF5: Track and record the career progression of employees who develop software skills through research activities, including transitions to academia, industry, and other sectors that depend on advanced computing expertise.}
\label{sec:CF5}

\subsection*{Description}
\begin{itemize}
  \item Motivation: Tracking career outcomes helps demonstrate the lasting value of software skills gained through research. By monitoring how employees transition into academic, industry, or technology-focused roles, institutions can show that software training is essential for research success and also enhances long-term employability. These data serve as evidence of impact beyond science, supporting internal advocacy for training programs, informing funding agencies about societal benefits, and emphasizing the importance of computing expertise in the broader job market. Even simple metrics, like the percentage of former staff in software-related roles—can showcase the extensive value of investing in research-based technical skills.
\end{itemize}

\subsection*{Class}
\begin{itemize}
  \item Cost, funding and return of investment
\end{itemize}

\subsection*{Actors}
\begin{itemize}
  \item home institute
\end{itemize}

\subsection*{Dependencies}
\begin{itemize}
  \item \hyperref[sec:SK2]{SK2}
  \item \hyperref[sec:SK1]{SK1}
  \item \hyperref[sec:SK3]{SK3}
  \item \hyperref[sec:SK5]{SK5}
  \item \hyperref[sec:SK4]{SK4}
\end{itemize}

\newpage

\section{CF6: Establish regular monitoring and review procedures to assess progress in data preservation and open science across the laboratory and its hosted experiments, ensuring accountability and ongoing improvement.}
\label{sec:CF6}

\subsection*{Description}
\begin{itemize}
  \item Motivation: Ongoing monitoring helps ensure that commitments to open science and data preservation lead to tangible, lasting action, not just policy statements. Regular reviews enable the laboratory to track progress, identify successful practices worth replicating, and address gaps before they become critical. These processes foster transparency and accountability to the research community, funding agencies, and institutional leadership, while also guiding resource allocation and long-term planning. They also promote knowledge sharing between experiments, support the adoption of best practices, and help align efforts with evolving community standards and technologies. Without consistent oversight, even the best policies risk being sidelined by competing priorities or a lack of follow-through.
\end{itemize}

\subsection*{Class}
\begin{itemize}
  \item Cost, funding and return of investment
\end{itemize}

\subsection*{Actors}
\begin{itemize}
  \item host laboratory
\end{itemize}

\subsection*{Dependencies}
\begin{itemize}
  \item \hyperref[sec:IR2]{IR2}
  \item \hyperref[sec:PM2]{PM2}
  \item \hyperref[sec:IR3]{IR3}
  \item \hyperref[sec:CF1]{CF1}
  \item \hyperref[sec:IR1]{IR1}
\end{itemize}

\newpage

\chapter{IC: International collaboration}

\section{IC1: Actively promote the exchange of experience and best practices in data preservation by engaging with the Data Preservation in High-Energy Physics (DPHEP) collaboration.}
\label{sec:IC1}

\subsection*{Description}
\begin{itemize}
  \item Motivation: Exchanging ideas and experiences from different stages of the experimental lifecycle, as well as current practices, is vital for understanding evolving needs and identifying effective solutions. Involving experts from both experiments and host laboratories ensures comprehensive knowledge transfer and helps anticipate future challenges in data preservation.
\end{itemize}

\subsection*{Class}
\begin{itemize}
  \item International collaboration
\end{itemize}

\subsection*{Actors}
\begin{itemize}
  \item host laboratory
  \item experiment management
  \item home institute
\end{itemize}

\newpage

\section{IC2: When serving on program committees or similar decision-making groups, actively advocate for dedicated sessions focused on the long-term preservation and reuse of research-quality data at major international conferences.}
\label{sec:IC2}

\subsection*{Description}
\begin{itemize}
  \item Motivation: Data preservation and open science topics are often grouped with sessions aimed at education and outreach, which are valuable but distinct. Dedicated sessions centered on preserving the full scientific value of research data would better raise awareness of the specific challenges and efforts required to ensure data remains accessible and usable for future scientific work.
\end{itemize}

\subsection*{Class}
\begin{itemize}
  \item International collaboration
\end{itemize}

\subsection*{Actors}
\begin{itemize}
  \item host laboratory
  \item experiment management
  \item WG leaders
  \item tool developers
\end{itemize}

\newpage

\appendix

\chapter{Audience Definitions}

\section{WG leaders}
Working Group leaders who coordinate specific scientific or technical subgroups within the collaboration, guiding research directions and ensuring progress on focused tasks.

\section{analysts}
Collaboration members who focus on interpreting experimental data, performing statistical analyses, and extracting scientific results from the collected information.

\section{data management}
The team or individuals in charge of organizing, storing, preserving, and providing access to the experiment’s data, ensuring its integrity and availability for analysis and future reference.

\section{experiment management}
The group of individuals responsible for overseeing the scientific, technical, and administrative aspects of the experimental collaboration, including planning, resource allocation, and day-to-day operations.

\section{funding agency}
An organization that provides financial support for high-energy physics (HEP) research and can establish requirements for open science, data preservation, and project accountability.

\section{home institute}
The primary institutional affiliation of individual researchers, such as universities, research institutes, or national laboratories, is responsible for supporting their involvement in open science and data stewardship activities.

\section{host laboratory}
A large-scale research facility that provides the physical infrastructure, computing resources, and institutional framework for conducting high-energy physics (HEP) experiments. It typically holds long-term responsibility for preserving experiment data and documentation.

\section{open data group}
A subgroup within the experiment dedicated to making selected experimental data publicly available.

\section{tool developers}
Individuals or teams responsible for creating and maintaining software or computing tools essential for the experiment’s data collection, processing, and analysis.

\newpage

\chapter{Glossary}

\section{Accessible}
Research data, software, and other digital objects are accessible when members of specified groups (e.g., the public, collaborators within an experiment) have sufficient permission to view and copy it, and when the resource and its metadata are retrievable via standardized, open protocols that may include authentication and authorization.

\section{Conditions data}
Data which describe the conditions of the detector such as alignment and calibration during data-taking periods. The conditions data are needed for reconstruction and analysis of data and are often stored in and made accessible via a conditions database.

\section{Container image}
A container image is a standalone executable package that includes everything needed to run an application together with its runtime dependencies. The container technology allows to encapsulate and version-control original computing environments in order to reuse them later. In contrast to virtual machines container images need to be executed on the same operating system and architecture that the image was built on.

\section{Continuous integration/ Continuous delivery (CI/CD)}
A software practice in which automated code building, testing, and deployment is done frequently.

\section{DPHEP}
Data Preservation in High-Energy Physics, a collaboration for data preservation and long-term analysis in high-energy physics.

\section{Data preservation}
The process of maintaining and safeguarding data so that it remains accessible and usable over the long term.

\section{Event-level data}
Data representing individual physics “events” that is typically used as the original input to a physics analysis. Event-level data are distinct from accumulated data such as histograms that are commonly preserved along with a publication.

\section{Experimental collaboration}
A structured partnership of researchers from multiple institutions working together on a specific HEP experiment. Collaborations operate under formal agreements and share collective responsibility for data management and research outputs.

\section{FAIR}
The FAIR principles are guidelines designed to improve how research data, software, and other digital objects are managed. FAIR stands for Findable, Accessible, Interoperable, and Reusable, with the understanding that these resources should be FAIR for both human and machine use.

\section{Findable}
Research data, software, and other digital objects are findable when they are assigned globally unique and persistent identifiers, described with rich metadata, and registered or indexed in a searchable resource. This ensures that both humans and computers can reliably locate and identify the objects.

\section{HEP}
High Energy Physics, a branch of physics that studies the fundamental particles and forces of the universe.

\section{Interoperable}
Research data, software, and other digital objects are interoperable if they are usable across different systems and software. This requires sufficient context for these resources in terms of metadata descriptions and the use of open formats and established standards for metadata and data.

\section{Legacy data}
The final, documented processing version of collected data and corresponding simulations from a given data-taking period, established as the reference for future analyses.

\section{Level 1 data}
As defined by DPHEP, data in support of publications (e.g. digitized plots or HepData records).

\section{Level 2 data}
As defined by DPHEP, data in simplified formats (e.g.for outreach or simple training).

\section{Level 3 data}
As defined by DPHEP, analysis-level event data and related software (for full scientific analysis based on existing reconstruction).

\section{Level 4 data}
As defined by DPHEP, raw data and related reconstruction and simulation software (full scientific potential of data).

\section{Long term}
A duration that extends beyond the lifespan of an experimental collaboration.

\section{Metadata}
Data that provide information about the data that the experiment was designed to measure. Metadata describe their context, characteristics, and structure.

\section{Open data}
Data that are openly accessible, exploitable, editable and shareable by anyone for any purpose. In general, open data are licensed under a specific open licence. 

\section{Open science}
The effort to make scientific research, its tools, processes and results accessible and available for the entire society. Open science encompasses open access to publications, open data, open source software and methods.

\section{Open source software}
Software in which the source code is made freely available for modification, distribution, and usage.

\section{Persistent identifier (DOI)}
A long-lasting unique reference to a resource such as a document, file, or dataset. A digital object identifier (DOI) is a persistent identifier for a digital resource.

\section{Reusable}
Research data, software, and other digital objects are reusable if it is specified sufficiently clear how they can be reused, including the use of machine-readable licenses and adherence to community standards.

\section{Software forge}
A web-based collaborative software platform for both developing and sharing code. This typically hosts several software repositories.

\section{Software license}
A software license is a legal instrument granted by copyright holders that governs the use, the modification and the redistribution of their software.

\section{Software repository}
A location where software code and related resources are stored and made accessible. It is often version-controlled.

\section{Supplementary data}
Data that support proper analysis of preserved data. Supplemental data may include calibrations, condition information, luminosity information, etc.

\section{Version control (Git)}
A software practice and system for tracking and managing changes in software code. Git is a widely-used version control system.

\section{Well-defined}
Well-defined computing environments fully specify the operating system and all software packages, including specific versions, needed to run a certain element of code.

\section{Workflow}
A sequence of - in the context of these recommendations - computational steps through which data are selected, processed, and analysed.

\newpage

\chapter{Useful Links}

This appendix provides a collection of useful links to resources, tools, and organizations relevant to high-energy physics and scientific computing.

\section{CDS}

CERN Document Server

\textbf{URL:} \url{https://repository.cern}

\section{CERNLIB}

A historical collection of software libraries and modules for high-energy physics.

\textbf{URL:} \url{https://cernlib.web.cern.ch/cernlib/}

\section{CODP}

CERN Open Data Portal.

\textbf{URL:} \url{https://opendata.cern.ch}

\section{CVMFS}

CERN VM File System, a network file system widely used in high-energy physics to distribute software and data across sites.

\textbf{URL:} \url{https://cernvm.cern.ch/fs/}

\section{DPHEP}

Data Preservation in High-Energy Physics, a collaboration for data preservation and long-term analysis in high-energy physics.

\textbf{URL:} \url{https://dphep.web.cern.ch}

\section{EVERSE}

European Virtual Institute for Research Software Excellence.

\textbf{URL:} \url{https://everse.software/}

\section{HDF5}

Hierarchical Data Format 5, a file format for storing large, complex, and heterogeneous datasets.

\textbf{URL:} \url{https://www.hdfgroup.org/solutions/hdf5/}

\section{HepData}

A repository for publication-related high-energy physics data.

\textbf{URL:} \url{https://www.hepdata.net/}

\section{HSF}

HEP Software Foundation, an organization that facilitates cooperation and common efforts in high-energy physics software and computing internationally.

\textbf{URL:} \url{https://hepsoftwarefoundation.org}

\section{OCI}

Open Container Initiative.

\textbf{URL:} \url{https://opencontainers.org/}

\section{OSI}

Open Source Initiative.

\textbf{URL:} \url{https://opensource.org/}

\section{REANA}

A platform for reproducible and reusable data analyses.

\textbf{URL:} \url{https://www.reana.io}

\section{ROOT}

An open-source data analysis framework used primarily by high-energy physics.

\textbf{URL:} \url{https://root.cern.ch}

\section{XRootD}

A software framework for data access.

\textbf{URL:} \url{https://xrootd.org/}

\section{Zenodo}

A general-purpose open repository for research-related content such as papers, datasets, software, and other digital artefacts. Each submission is assigned a DOI.

\textbf{URL:} \url{https://zenodo.org}

\newpage

\end{document}